\documentclass[12pt]{article}
\usepackage{amsfonts}
\usepackage{amsmath}
\usepackage{amssymb}
\usepackage{amsthm}

\usepackage[utf8]{inputenc}
\usepackage[margin=2cm]{geometry}
\usepackage{geometry}
\usepackage{hyperref}
\usepackage{graphicx}
\usepackage{amsmath}
\usepackage{cite}
\usepackage{hyperref}
\usepackage{tensor}
\usepackage{cases}
\usepackage{appendix}
\usepackage{multirow}
\usepackage{booktabs}
\usepackage{color}
\usepackage[font=small]{subfig}
\usepackage[skip=8pt,labelfont={bf,sc},textfont=small]{caption}

\def\[#1\]{\begin{align}#1\end{align}}
\newcommand{\imineq}[2]{\vcenter{\hbox{\includegraphics[height=#2ex]{#1}}}}
\def \nn {\nonumber}

\def \dd{\mathrm{d}}

\def \e {\epsilon}

\def \ra{\rangle}
\def \la{\langle}

\def \mo{\mathcal{O}}

\def \tr{\text{tr}}
\def \T{\mathcal{T}}

\def \LHP{\text{LHP}}
\def \UHP{\text{UHP}}
\def \AdS{\text{AdS}}
\def \of{\text{of}}
\def \Arctanh{\text{Arctanh}}
\begin{document}
\begin{titlepage}
%\begin{flushright}
%TIT/HEP-6XX \\
%mm,  2016
%\end{flushright}
\vspace{0.5cm}
\begin{center}
{\Large \bf {Pseudo entropy of primary operators in $T$$\bar{T}$/$J\bar{T}$-deformed CFTs}}
\lineskip .75em
\vskip 2.5cm
{Song He$^{a,c,}$\footnote{hesong@jlu.edu.cn}, Jie Yang$^{b,}$\footnote{yangjie@cnu.edu.cn},  Yu-Xuan Zhang$^{a,}$\footnote{yuxuanz20@mails.jlu.edu.cn}, Zi-Xuan Zhao$^{a,}$\footnote{zhaozixuan@cnu.edu.cn}}
\vskip 2.5em
{\normalsize\it  
$^{a}$Center for Theoretical Physics and College of Physics, Jilin University,\\ Changchun 130012, People's Republic of China\\
$^{b}$School of Mathematical Sciences, Capital Normal University,\\
Beijing 100048, People's Republic of China\\
$^{c}$Max Planck Institute for Gravitational Physics (Albert Einstein Institute),\\
Am M\"uhlenberg 1, 14476 Golm, Germany}\\
\vskip 1.0em
\vskip 3.0em
\end{center}
\begin{abstract}
%This work calculates the first-order corrections to the  pseudo-R\'enyi entropy of primary operators in the $T\bar{T}/J\bar{T}$-deformed CFTs. 

%In this work, we investigate the time evolution of the pseudo-(R\'enyi) entropy perturbatively after local operator quenches in 2D CFTs with $T\bar T/J\bar T$-deformation. We focus on states prepared by acting primary operators located at different spatial locations on the vacuum. The corrections of the $k^{\rm th}$ pseudo-R\'enyi entropy at late times are obtained in terms of the perturbative CFT approach. The corresponding corrections of pseudo entropy at late times are obtained by taking the limit of $k \to 1.$
In this work, we investigate the time evolution of the pseudo-(R\'enyi) entropy after local primary operator quenches in 2D CFTs with $T\bar T/J\bar T$-deformation. Using perturbation theory, we analyze the corrections to the second pseudo-R\'enyi entropy at the late time, which exhibit a universal form, while its early-time behavior is model-dependent. Moreover, we uncover nontrivial time-dependent effects arising from the first-order deformation of the $k^{\rm th}$ pseudo-R\'enyi entropy at the late time. Additionally, drawing inspiration from the gravitational side, specifically the gluing of two cutoff AdS geometries, we investigate the $k^{\rm th}$ pseudo-R\'enyi entropy for vacuum states characterized by distinct $T\bar{T}$-deformation parameters, as well as for primary states acting on different deformed vacuum states. Our findings reveal additional corrections compared to the results of pseudo-R\'enyi entropy for globally deformed vacuum states. %These additional corrections may introduce new boundary conditions when we perform the gluing procedure for two cutoff AdS geometries.
%We obtain a universal perturbative correction to the second pseudo-R\'enyi entropy in the late-time limit, while the early-time behavior is model-depended. We also provide results for the first-order corrections to the $k^{\rm th}$ R\'enyi entropy of excited states in $T\bar{T}/J\bar{T}$-deformed two-dimensional CFTs at both the late and early time. Thus we demonstrate nontrivial time-dependent behavior for the first-order deformation to the $k^{\rm th}$ pseudo-R\'enyi entropy.

%Additionally, inspired by gluing two cutoff AdS geometries on the gravity side we evaluate the $k^{\rm th}$ pseudo-R\'enyi entropy for vacuum states and primary states with different $T\bar{T}$-deformation parameters.

%In this paper, we calculate the first-order correction to the pseudo-entropy of primary operators in the  $T\bar{T}/J\bar{T}$-deformed CFTs. The resulting first-order correction to the second pseudo-R\'enyi entropy in the late time limit has a universal form, while the early time behavior depends on the theory details. Further, the $k$-th R\'enyi entropy of the excited states has been obtained at the late and early time in the $T\bar{T}/J\bar{T}$-deformed two-dimensional CFTs. We find nontrivial time-dependent behavior of the first-order deformation to the entanglement entropy, the $k$-th pseudo-R\'enyi entropy, and pseudo-entropy. In particular, inspired by gluing two different cutoff AdS geometries on the gravity side, we evaluate the $k$-th pseudo-R\'enyi entropy for vacuum states with different $T\bar{T}$ deformation parameters.

\end{abstract}
\end{titlepage}

\baselineskip=0.7cm

\tableofcontents
\section{Introduction}
The AdS/CFT correspondence\cite{Maldacena:1997re, Gubser:1998bc, Witten:1998qj} provides a holographic description of a quantum gravity theory in an asymptotically anti-de Sitter (AdS) spacetime with a conformal field theory (CFT) at its asymptotic boundary. Its discovery has inspired significant exploration regarding quantum information theory in the high-energy physics community in recent years, covering a range of topics such as quantum entanglement\cite{Casini:2004bw, Calabrese:2004eu, Kitaev:2005dm, Casini:2016fgb, Nishioka:2018khk, Witten:2018zxz, Casini:2022rlv}, the emergence of geometry \cite{VanRaamsdonk:2010pw, Maldacena:2013xja, Rangamani:2016dms}, and the black hole information paradox \cite{Hawking:1976ra, Mathur:2009hf, Almheiri:2012rt, Penington:2019npb, Almheiri:2019psf}.

Recently, there has been a surge of interest in 2D CFTs deformed by $T\bar{T}/J\bar{T}$ operator\cite{Smirnov:2016lqw,Cavaglia:2016oda,Guica:2017lia}. Though the deformations are irrelevant in the sense of the renormalization group, owing to the integrability\cite{Smirnov:2016lqw, Guica:2017lia} and holographic duality \cite{McGough:2016lol, Guica:2019nzm, Kraus:2018xrn, Chakraborty:2018vja, Bzowski:2018pcy}, they have been extensively investigated in recent years\cite{Dubovsky:2017cnj, Cardy:2018sdv, Aharony:2018vux, Bonelli:2018kik, Datta:2018thy, Aharony:2018bad, Conti:2018tca, Chang:2018dge, Conti:2018jho, Jiang:2023ffu, Jeong:2022jmp, Gross:2019uxi, Geng:2019yxo, He:2019ahx, Asrat:2020uib, Ouyang:2020rpq, Li:2020pwa, Ebert:2020tuy, Medenjak:2020ppv, Ebert:2022xfh, Li:2020zjb, He:2020qcs, Jiang:2019trm, Chakraborty:2020udr, He:2020hhm, Aharony:2023dod, He:2023hoj,Basu:2023bov,Jeong:2019ylz,He:2022ryk}. The $T\bar T/J\bar{T}$-deformation triggers a flow of the original theory along a trajectory in the space of field theory.
%action corresponds to a trajectory in the space of field theory.
For the $T\bar T$-deformation, the trajectory of the deformed action satisfies the following condition:
\[
\frac{d S(\lambda)}{d\lambda}=\int d^2 z \sqrt{g}(T\bar T)_\lambda,
\]
where $\lambda$ represents the coupling constant of the $T\bar{T}$ operator. $S(\lambda=0)$ corresponds to the action of the undeformed theory. In the flat space, the $T\bar{T}$ operator can be expressed as
\[
T\bar{T}=-\frac{1}{4}\text{det}(T_{\mu\nu})=T_{zz}T_{\bar{z}\bar{z}}-T_{z\bar{z}}T_{z\bar{z}},
\]
with $T=T_{zz}$ and $\bar{T}=T_{\bar{z}\bar{z}}$. In this paper,  we mainly focus on the perturbation theory of $S(\lambda)$. The $T\bar T$-deformed action, up to the first order of $\lambda$, is given by
\[
S(\lambda)=S(\lambda=0)+\lambda\int d^2 z \sqrt{g}(T\bar T)_{\lambda=0}+\mo(\lambda^2),
\]
and we refer to $(T\bar T)_{\lambda=0}$ as simply $T\bar T$. We treat the $J\bar T$-deformation in the same manner.
%The $T\bar{T}$-deformation is determined by the holomorphic and anti-holomorphic part of the stress tensor and triggers a flow in CFTs. 
%This $T\bar{T}$ operator is well-defined in any translation invariant 2D CFTs. 
%By virtue of the property that $T\bar{T}$ operator is well-defined in any 2D CFTs, specific quantities such as the S-matrix and the spectrum of any $T\bar{T}$-deformed CFTs, can be calculated entirely based on the data of the undeformed theory\cite{Smirnov:2016lqw, Cavaglia:2016oda}. Similar computation has been considered for $J\bar{T}$-deformed CFTs\cite{He:2020qcs,Jiang:2019trm,Chakraborty:2020udr,He:2020hhm}.

Pseudo entropy is a generalization of the entanglement entropy proposed via the AdS/CFT correspondence and post-selection \cite{Nakata:2020luh}. %which is a valuable quantity to characterize both the vacuum and excited states. Its usefulness has been explored in recent works \cite{Mollabashi:2020yie, Mollabashi:2021xsd, Guo:2022sfl, He:2023eap}.
Specifically,  the pseudo entropy is a two-state vector version of the entanglement entropy, defined as follows. Let $|\psi\rangle$ and $|\varphi\rangle$ be two non-orthogonal states in the Hilbert space $\mathcal{H}_{\mathcal{S}}$ of a composite quantum system $\mathcal{S}=A\cup B$. We initially construct an operator acting on $\mathcal{H}_{\mathcal{S}}$ which is called the \textit{transition matrix} \cite{Nakata:2020luh,Guo:2022jzs}
\[
\T^{\psi|\varphi}\equiv\frac{|\psi\ra\la\varphi|}{\la\varphi|\psi\ra}=\frac{\rho_\psi\rho_\varphi}{\tr[\rho_\psi\rho_\varphi]}.\label{transitionmatrix}
\]
 We then define the so-called \textit{reduced transition matrix} $\mathcal{T}_A^{\psi|\varphi}$ for subsystem $A$  by tracing out the degrees of freedom in the subsystem $B$ from the transition matrix \eqref{transitionmatrix}, that is, $\mathcal{T}_A^{\psi|\varphi}=\text{Tr}_B[\T^{\psi|\varphi}]$. %In the case of a bipartite system $\mathcal{S}=A\cup B$, where $A$ and $B$ are non-overlapping subsystems, the reduced transition matrix is a square matrix of size $d_A\times d_A$, where $d_A$ is the dimension of the Hilbert space of subsystem $A$. 
 The pseudo entropy of $A$  is then defined by the von Neumann entropy of  $\mathcal{T}_A^{\psi|\varphi}$, i.e.,
 \[
 S_{A}=-\tr\big[\mathcal{T}_A^{\psi|\varphi}\log\mathcal{T}_A^{\psi|\varphi}\big].\label{pedef}
 \]
Furthermore, we can introduce the so-called \textit{pseudo-R\'enyi entropy} $S_A^{(k)}$, a generalization of the R\'enyi entropy %which is defined as 
\[S_A^{(k)}=\frac{1}{1-k}\log\text{tr}[(\rho_A)^k],\label{renyidef}
\] 
utilizing the transition matrix.
%where $\rho_A$ is the reduced density matrix of subsystem $A$.
The pseudo-R\'enyi entropy is  defined by replacing the reduced density matrix in the R\'enyi entropy \eqref{renyidef} with the reduced transition matrix $\mathcal{T}_A^{\psi|\varphi}$,
\[
S_A^{(k)}=\frac{1}{1-k}\log\tr\big[\big(\mathcal{T}_A^{\psi|\varphi}\big)^k\big].\label{pseudorenyi}
\] 
Note that the pseudo-R\'enyi entropy \eqref{pseudorenyi} gives the pseudo entropy \eqref{pedef} in the limit of $k\rightarrow 1$.

It has been shown that pseudo entropy, like entanglement entropy, can be used to quantify the topological contributions of entanglement  of excited states in
two-dimensional rational CFTs \cite{ Nishioka:2021cxe,Guo:2022sfl,He:2023eap}\footnote{Please refer to \cite{Mollabashi:2020yie,Camilo:2021dtt,Mollabashi:2021xsd,Goto:2021kln,Miyaji:2021lcq,Akal:2021dqt
,Mukherjee:2022jac,Ishiyama:2022odv,Doi:2022iyj,Li:2022tsv,Doi:2023zaf,Chen:2023gnh,Narayan:2023ebn
, Jiang:2023loq, Chu:2023zah,He:2023ubi,Parzygnat:2023avh,Chen:2023eic,Guo:2023aio} for other recent development of pseudo entropy.}.
%In the course of computing the pseudo entropy, the pseudo-R\'enyi entropy has been extensively studied in\cite{Mollabashi:2021xsd,Nishioka:2021cxe, Goto:2021kln,Mukherjee:2022jac, Guo:2022sfl, He:2023eap}. 
The behavior of the entanglement entropy in the deformed CFTs, especially in the context of the $T\bar{T}$- and $J\bar{T}$-deformations, has been extensively studied in recent years\cite{Chen:2018eqk,  He:2023xnb, He:2022xkh, Donnelly:2018bef, Ashkenazi:2023fcn}. However, little is known about the pseudo-R\'enyi entropy in $T\bar{T}/J\bar{T}$-deformed CFTs.
Another motivation for investigating pseudo-R\'enyi entropy in $T\bar{T}/J\bar{T}$-deformed CFTs comes from the holography. It is thought that pseudo entropy has its holographic dual, and the holographic dual of the positive sign $T\bar{T}$-deformed two-dimensional holographic CFT was proposed to be $\AdS_3$ gravity with a finite radius cut-off. So the properties of the pseudo-R\'enyi entropy in the deformed CFTs would provide insight into the structure of the deformed theories and their relationship to the undeformed CFTs. Therefore, investigating the pseudo-R\'enyi entropy in the deformed CFTs is an interesting and important research direction. 

This paper aims to investigate the pseudo-R\'enyi entropy in the $T\bar{T}$/$J\bar{T}$-deformed CFTs. Our inquiry can be traced back to the investigations on the R\'enyi entropy of vacuum states\cite{Chen:2018eqk} and locally excited states\cite{He:2019vzf} in the $T\bar{T}$/$J\bar{T}$-deformed theories. It has been observed that, with the first-order perturbation of the deformation, the second R\'enyi entanglement entropy of locally excited states acquires a nontrivial time dependence. It would be captivating to investigate the correction of the pseudo-R\'enyi entropy in the deformed CFTs. Besides, inspired by gluing together two cutoff AdS geometries and the cutoff AdS/$T\bar{T}$-deformed-CFT(cAdS/dCFT) correspondence, we are interested in exploring the pseudo-R\'enyi entropy of two states with two different deformation parameters.

The structure of this paper is as follows. In section \ref{setup}, we briefly overview the $T\bar{T}$-deformation of the correlation functions and the replica method for locally excited states in 2D CFTs. In section \ref{section3}, we concentrate on the second pseudo-R\'enyi entropy of the $T\bar{T}$-deformed theories. In section \ref{section4}, we investigate the excess of the $k^{\rm th}$ R\'enyi entropy of locally excited states and the $k^{\rm th}$ pseudo-R\'enyi entropy of locally excited states in $T\bar{T}$-deformed CFTs. In section \ref{section5}, we compute the $k^{\rm th}$ pseudo-R\'enyi entropy of two vacuum states and two primary states with different deformation parameters in $T\bar{T}$-deformed CFTs. In section \ref{section6}, we extend the analysis from sections \ref{section3} and \ref{section4} to the $J\bar{T}$-deformed theory. Finally, in section \ref{conclusion}, we provide concluding remarks and discuss possible future research directions. Some technical details of the calculations are presented in the appendices.

\section{Setup}\label{setup}

\subsection{The $T\bar{T}$-deformation  of the correlation  functions}
In this subsection, we briefly overview the $T\bar{T}$-deformation in 2D CFTs and its effect on the correlation functions in \cite{He:2019vzf}.

%The $T\bar T$-deformed action corresponds to a trajectory in the space of field theory, where the trajectory satisfies the following condition:
%\[
%\frac{d S(\lambda)}{d\lambda}=\int d^2 z %\sqrt{g}(T\bar T)_\lambda,
%\]
%where $\lambda$ represents the coupling constant of the $T\bar{T}$ operator. The action $S(\lambda=0)$ corresponds to the undeformed CFT on a flat space. In this flat space, the $T\bar{T}$ operator can be expressed as
%\[
%T\bar{T}=\text{det}%(T_{\mu\nu})=T_{zz}T_{\bar{z}\bar{z}}-%T_{z\bar{z}}T_{z\bar{z}},
%\]
%where $T=T_{zz}$ and $\bar{T}=T_{\bar{z}\bar{z}}$. 
%we focus on the perturbation theory of $S(\lambda)$ and expand it perturbatively as
%\[
%S(\lambda)=S(\lambda=0)+\lambda\int d^2 z \sqrt{g}(T\bar T)_{\lambda=0}+O(\lambda^2).
%\]
%Henceforth in this paper, we refer to $(T\bar T)_{\lambda=0}$ as simply $T\bar T$.

By using the Ward identity, the first-order correction to the $k$-point correlation function of primary operators in $T\bar{T}$-deformed CFTs can be expressed as $\langle \mo_1(z_1,\bar{z}_1)\cdots\mo_k(z_k,\bar{z}_k)\rangle_\lambda$, where $\lambda$ is the coupling constant of the $T\bar{T}$-deformation. The expression for the first-order correction is as follows:
\[
 &\la\mo_{1}(z_1,\bar{z}_1)\mo_{2}(z_2,\bar{z}_2)\dots\mo_{k}(z_k,\bar{z}_k)\ra_{\lambda}\nonumber\\
 =&\lambda \int d^2 z\left\langle T \bar{T}(z, \bar{z}) \mo_1\left(z_1, \bar{z}_1\right) \mo_2\left(z_2, \bar{z}_2\right) \cdots \mo_k\left(z_k, \bar{z}_k\right)\right\rangle\nn\\
=&\lambda\int d^2z\left(\sum\limits^{k}_{i=1}\left(\frac{h_i}{(z-z_i)^2}+\frac{\partial_{z_i}}{z-z_i}\right)\right)\left(\sum\limits^{k}_{i=1}\left(\frac{\bar{h}_i}{(\bar{z}-\bar{z}_i)^2}+\frac{\partial_{\bar{z}_i}}{\bar{z}-\bar{z}_i}\right)\right)\nonumber\\
&\times \la\mo_{1}(z_1,\bar{z}_1)\mo_{2}(z_2,\bar{z}_2)\dots\mo_{k}(z_k,\bar{z}_k)\ra. \label{equ1.1}
\]
To obtain the result, we take advantage of the fact that any correlation function involving $T_{z\bar{z}}$ vanishes in a 2D CFT. Specifically, we begin with the two-point correlation function of primary operators $\mo_1$ and $\mo_2$ with conformal weights $(h, \bar{h})$. The two-point function of $\mo_1$ and $\mo_2$ can be written as
\[
\la\mo_{1}(z_1,\bar{z}_1)\mo_{2}(z_2,\bar{z}_2)\ra=\frac{C_{12}}{z_{12}^{2h}\bar{z}_{12}^{2\bar{h}}}, \label{equ1.2}
\]
where $z_{ij}=z_i-z_j$, $\bar z_{ij}=\bar z_i-\bar z_j$. The Ward identity of the two-point function gives us the following relation:
\[
&\la T(z)\bar{T}(\bar{z})\mo_{1}(z_1,\bar{z}_1)\mo_{2}(z_2,\bar{z}_2)\ra \nonumber\\
=&\bigg( \frac{h\bar{h}z_{12}^2 \bar{z}_{12}^2}{(z-z_1)^2(z-z_2)^2(\bar{z}-\bar{z}_1)^2(\bar{z}-\bar{z}_2)^2}-\frac{4\pi \bar{h}}{(z-z_1)(\bar{z}_1-\bar{z}_2)}\delta^{(2)}(z-z_1) \nonumber\\
& -\frac{2\pi \bar{h}}{(z-z_2)(\bar{z}_1-\bar{z}_2)}\delta ^{(2)}(z-z_2)\bigg)\la\mo_{1}(z_1,\bar{z}_1)\mo_{2}(z_2,\bar{z}_2)\ra. \label{equ1.3}
\]
The appearance of the delta function arises from the action of $\partial_{z_i}$ on terms such as $(\bar{z}-\bar{z}_i)^{-1}$, which leads to a non-dynamical contribution. After the integration over spacetime, this contribution gives rise to a UV divergence. To simplify the following analysis, we omit these terms. We can obtain the first-order correction of the two-point function resulting from the $T\bar{T}$-deformation using dimension regularization
\[
\la \mo_{1}(z_1,\bar{z}_1)\mo_{2}(z_2,\bar{z}_2)\ra_\lambda
=\lambda h\bar{h}\frac{4\pi}{\left|z_{12}\right|^2}\left(\frac{4}{\epsilon}+2\log(\left|z_{12}\right|^2)+2\log\pi+2\gamma-5\right)\la\mo_{1}(z_1,\bar{z}_1)\mo_{2}(z_2,\bar{z}_2)\ra. \label{equ1.4}
\]
Next, we compute the four-point function within the framework of $T\bar{T}$-deformed CFTs. Specifically, the four-point function of primary operators can be written as follows:
\[
\la \mo^{\dagger}_{1}(z_1,\bar{z}_1)\mo_{2}(z_2,\bar{z}_2)\mo^{\dagger}_{1}(z_3,\bar{z}_3)\mo_{2}(z_4,\bar{z}_4)\ra=\frac{G(\eta,\bar{\eta})}{z_{13}^{2h}z_{23}^{2h}\bar{z}_{13}^{2\bar{h}}\bar{z}_{24}^{2\bar{h}}}, \label{equ1.5}
\]
with the cross ratios
\[
\eta:=\frac{z_{12}z_{34}}{z_{13}z_{24}}, \quad\bar{\eta}:=\frac{\bar{z}_{12}\bar{z}_{34}}{\bar{z}_{13}\bar{z}_{24}}, \label{equ1.6}
\]
and $G(\eta,\bar\eta)$ is the conformal block which is a function of cross ratios. 

The first-order correction to this four-point function due to the $T\bar{T}$-deformation is given by: %{\color{red}has the following sentence missed  $T\bar T$}
\[
&\la \mo^{\dagger}_{1}(z_1,\bar{z}_1)\mo_{2}(z_2,\bar{z}_2)\mo^{\dagger}_{1}(z_3,\bar{z}_3)\mo_{2}(z_4,\bar{z}_4)\ra_{\lambda}\nonumber\\
=& \lambda\int d^2 z \left\{\left( \frac{h z_{13}^2}{(z-z_1)^2(z-z_3)^2}+\frac{h z_{24}^2}{(z-z_2)^2(z-z_4)^2}+\frac{z_{23}z_{14}}{\prod_{i=1}^{4}(z-z_i)}\frac{\eta G(\eta,\bar{\eta})}{G(\eta,\bar{\eta})}\right)\right.\nonumber\\
&\times\left( \frac{\bar{h} \bar{z}_{13}^2}{(\bar{z}-\bar{z}_1)^2(\bar{z}-\bar{z}_3)^2}+\frac{\bar{h} \bar{z}_{24}^2}{(\bar{z}-\bar{z}_2)^2(\bar{z}-\bar{z}_4)^2}+\frac{\bar{z}_{23}\bar{z}_{14}}{\prod_{i=1}^{4}(\bar{z}-\bar{z}_i)}\frac{\bar{\eta} G(\eta,\bar{\eta})}{G(\eta,\bar{\eta})}\right)\nonumber\\
&-\eta\bar{\eta}\frac{z_{23}z_{14}\bar{z}_{23}\bar{z}_{14}}{\prod_{i=1}^4 (z-z_i)(\bar{z}-\bar{z}_i)}\frac{\partial_{\eta}G(\eta,\bar{\eta})\partial_{\bar{\eta}}G(\eta,\bar{\eta})}{G(\eta,\bar{\eta})^2}+\eta\bar{\eta}\frac{z_{23}z_{14}\bar{z}_{23}\bar{z}_{14}}{\prod_{i=1}^4 (z-z_i)(\bar{z}-\bar{z}_i)}\frac{\partial_{\eta}\partial_{\bar{\eta}}G(\eta,\bar{\eta})}{G(\eta,\bar{\eta})}\nonumber\\
&-\frac{4\pi\bar{h}\bar{z}_{13}^2}{(z-z_1)(\bar{z}-\bar{z}_1)(\bar{z}-\bar{z}_3)^2}\delta^{(2)}(z-z_1)-\frac{4\pi\bar{h}\bar{z}_{13}^2}{(z-z_3)(\bar{z}-\bar{z}_3)(\bar{z}-\bar{z}_1)^2}\delta^{(2)}(z-z_3)\nonumber\\
&-\frac{4\pi\bar{h}\bar{z}_{24}^2}{(z-z_2)(\bar{z}-\bar{z}_2)(\bar{z}-\bar{z}_4)^2}\delta^{(2)}(z-z_2)-\frac{4\pi\bar{h}\bar{z}_{24}^2}{(z-z_4)(\bar{z}-\bar{z}_4)(\bar{z}-\bar{z}_2)^2}\delta^{(2)}(z-z_2)\nonumber\\
&-\frac{\bar{z}_{12}\bar{z}_{34}\partial_{\bar{\eta}}G(\eta,\bar{\eta})}{G(\eta,\bar{\eta})}\bigg[\frac{2\pi \delta^{(2)}(z-z_1)}{(z-z_1)(\bar{z}-\bar{z}_2)(\bar{z}-\bar{z}_3)(\bar{z}-\bar{z}_4)}+\frac{2\pi \delta^{(2)}(z-z_2)}{(z-z_2)(\bar{z}-\bar{z}_1)(\bar{z}-\bar{z}_3)(\bar{z}-\bar{z}_4)}\nonumber\\
&\left.+\frac{2\pi \delta^{(2)}(z-z_3)}{(z-z_3)(\bar{z}-\bar{z}_1)(\bar{z}-\bar{z}_2)(\bar{z}-\bar{z}_4)}+\frac{2\pi \delta^{(2)}(z-z_4)}{(z-z_4)(\bar{z}-\bar{z}_1)(\bar{z}-\bar{z}_2)(\bar{z}-\bar{z}_3)}\bigg]\right\}\nonumber\\
&\times\la \mo^{\dagger}_{1}(z_1,\bar{z}_1)\mo_{2}(z_2,\bar{z}_2)\mo^{\dagger}_{1}(z_3,\bar{z}_3)\mo_{2}(z_4,\bar{z}_4)\ra. \label{equ1.7}
\]

\subsection{The pseudo entropy and the replica method}
%In this subsection, we provide a brief review of the replica method to construct the pseudo-R\'enyi entropy with locally excited states in rational conformal field theories (RCFTs). We begin by considering a RCFT on a plane with a vacuum state represented by $|\Omega\ra$. We create two locally excited states by applying the primary operators $\mo(x_1)$ and $\mo^\dagger(x_2)$ to the vacuum state and then use them to generate a real-time evolved transition matrix denoted by $\mathcal{T}^{1|2}(t)$,
This subsection presents a concise overview of the replica method for constructing the pseudo-R\'enyi entropy with locally excited states in rational conformal field theories (RCFTs). We start by considering a RCFT defined on a plane, where the vacuum state is denoted as $|\Omega\ra$. By applying the primary operators $\mo(x_1)$ and $\mo^\dagger(x_2)$ to the vacuum state, we create two locally excited states
\[
|\psi_1\rangle\equiv e^{-\epsilon H}\mo(x_1)|\Omega\rangle,\quad
|\psi_2\rangle\equiv e^{-\epsilon H}\mo^{\dagger}(x_2)|\Omega\rangle,
\]
where the infinitesimal parameter  $\epsilon$ is introduced to suppress the high energy modes \cite{Calabrese:2005in}.
Subsequently, we utilize these states to construct a real-time evolved transition matrix denoted as $\mathcal{T}^{1|2}(t)$:
\begin{align}
\mathcal{T}^{1|2}(t)\equiv e^{-iH t}\frac{|\psi_1\rangle\langle\psi_2|}{\langle\psi_2|\psi_1\rangle}e^{iHt}.\label{equ1.8}
\end{align}
  To obtain the reduced transition matrix of subsystem $A$ at time $t$, we trace out the degrees of freedom of $A^c$ (the complement of $A$) from the transition matrix $\mathcal{T}^{1|2}(t)$, giving us $\mathcal{T}^{1|2}_A(t)=\tr{A^c}[\T^{1|2}(t)]$. 
 
 It turns out that the excess of the $k^{\rm th}$ pseudo-R\'enyi entropy of $A$ with respect to the ground state, defined as $\Delta S^{(k)}(\T_{A}^{1|2}(t)):=S^{(k)}(\T_A^{1|2}(t))-S^{(k)}\big(\tr_{A^c}\big[|\Omega\rangle\langle\Omega|\big]\big)$, is of the form \cite{Guo:2022sfl}
 \[
\Delta S^{(k)}(\mathcal{T}_A^{1|2}(t))=\frac{1}{1-k}\big[&\log\langle \prod_{j=1}^{k} \mo(w_{2j-1},\bar{w}_{2j-1})\mo^{\dagger}(w_{2j},\bar{w}_{2j})\rangle_{\Sigma_k}\nn\\
-&k\log \langle \mo(w_1,\bar{w}_1)\mo^{\dagger}(w_2,\bar{w}_2) \rangle_{\Sigma_1}\big], \label{equ1.9}
\]
according to the replica method. In Eq.$\eqref{equ1.9}$, $\Sigma_k$ denotes an  $k$-sheet Riemann surface with cuts on each copy corresponding to $A$, and  $(w_{2j-1},\bar{w}_{2j-1})$ and $(w_{2j},\bar{w}_{2j})$  are coordinates on the $j^{\rm th}$-sheet surface. The term in the first line of Eq.(\ref{equ1.9}) is given by a $2k$-point correlation function on $\Sigma_k$, and the one in the second line is calculated from a two-point function on $\Sigma_1$. The coordinates are chosen as 
\[
w_{2j-1}&=x_1+t-i\epsilon,\quad w_{2j}=x_2+t+i\epsilon,\nn\\
\bar{w}_{2j-1}&=x_1-t+i\epsilon,\quad\bar{w}_{2j}=x_2-t-i\epsilon.\quad (j=1,2,...,k)\label{equ1.10}
\]
The $2k$-point correlation function on $\Sigma_k$ in Eq.\eqref{equ1.9} can be evaluated with the help of a conformal mapping 
of $\Sigma_k$ to the complex plane $\Sigma_1$. For convenience, the subsystem is chosen such that $A=[0,\infty)$ hereafter. We can then map $\Sigma_k$ to $\Sigma_1$ using the simple conformal mapping 
\[
w=z^{k}\label{equ1.11}.
\]
We subsequently utilize the mapping given by Eq. \eqref{equ1.11} to express the $2k$-point function on $\Sigma_k$,
\[
&\la\mo(w_{1},\bar{w}_{1})\mo^{\dagger}(w_{2},\bar{w}_{2})\dots\mo(w_{2k-1},\bar{w}_{2k-1})\mo^{\dagger}(w_{2k},\bar{w}_{2k})\ra_{\Sigma_k} \nonumber \\
=& \left(\prod\limits_{j=1}^{2k}\left|\frac{d w_j}{d z_j}\right|^{-2 h}\right)\la\mo(w_{1},\bar{w}_{1})\mo^{\dagger}(w_{2},\bar{w}_{2})\dots\mo(w_{2k-1},\bar{w}_{2k-1})\mo^{\dagger}(w_{2k},\bar{w}_{2k})\ra_{\Sigma_1}\label{equ1.12}.
\]
The $2k$ points $z_1, z_2,...,z_{2k}$ in the $z$-coordinates are given by
\[
z_{2j+1}=&\text{e}^{2\pi i\frac{j+1/2}{k}}(-x_1-t+i\e)^{\frac{1}{k}},\quad \bar z_{2j+1}=\text{e}^{-2\pi i\frac{j+1/2}{k}}(-x_1+t-i\e)^{\frac{1}{k}},\nn\\
z_{2j+2}=&\text{e}^{2\pi i\frac{j+1/2}{k}}(-x_2-t-i\e)^{\frac{1}{k}},\quad \bar z_{2j+2}=\text{e}^{-2\pi i\frac{j+1/2}{k}}(-x_2+t+i\e)^{\frac{1}{k}},\quad (j=0,...,k-1).\label{A=INFzcoordin}
\]
\section{Second pseudo-R\'enyi entropy in $T\bar{T}$-deformed CFTs}\label{section3}

%\textcolor{blue}{The deformation of second pseudo-R\'enyi entropy can be witten as--(introducing the notation)}
The late-time excess of pseudo-R\'enyi entropy of locally excited operators has a universal behavior: the logarithmic quantum dimension of the operator, while the early-time one does not. Investigating the late-time behavior of the excess of pseudo-R\'enyi entropy and its behavior in the early time in $T\bar{T}$-deformed CFTs is intriguing.

In \cite{Chen:2018eqk}, the authors discussed R\'enyi entropy 
in the $T\bar{T}$-deformed CFTs for vacuum theory. It is shown that up to the first order of the deformation parameter $\lambda$,
\[
\Delta S_{A,\lambda}^{(k)}=\frac{-k\lambda}{1-k}\int_{\Sigma_1}[\langle T \bar{T}\rangle_{\Sigma_k}-\langle T \bar{T}\rangle_{\Sigma_1}]+\mo(\lambda^2),
\]
where we denote $\Delta S_{A,\lambda}^{(k)}$ as the correction to the undeformed R\'enyi entropy.
%\[\int_{\Sigma_n}\langle T \bar{T}\rangle_{\Sigma_n}=n \int_{\Sigma_1}\langle T \bar{T}\rangle_{\Sigma_n}.\]

Now let us consider second pseudo-R\'enyi entropy in $T\bar{T}$-deformed CFTs
\[
&-\left( \Delta S_{A,0}^{(2)}+\Delta S_{A,\lambda}^{(2)}\right)\nonumber\\
%-\left( \Delta S_{A}^{(2)}\right)_{T\bar T}\nonumber\\
=& \log \bigg( \la\mo(w_{1},\bar{w}_{1})\mo^{\dagger}(w_{2},\bar{w}_{2})\mo(w_{3},\bar{w}_{3})\mo^{\dagger}(w_{4},\bar{w}_{4})\ra_{\Sigma_2}\nonumber\\
&+2\lambda \int d^2 w \la T\bar{T}(w,\bar{w})\mo(w_{1},\bar{w}_{1})\mo^{\dagger}(w_{2},\bar{w}_{2})\mo(w_{3},\bar{w}_{3})\mo^{\dagger}(w_{4},\bar{w}_{4})\ra_{\Sigma_2}\bigg)\nonumber\\
&-2\log \bigg( \la \mo (w_1,\bar{w}_1)\mo^{\dagger}(w_2,\bar{w}_2)\ra_{\Sigma_1}+\lambda \int d^2 w \la T\bar{T}(w,\bar{w}) \mo (w_1,\bar{w}_1)\mo^{\dagger}(w_2,\bar{w}_2)\ra_{\Sigma_1} \bigg). \label{equ2.1}
\]
The derivation of \eqref{equ2.1} is a simplified version of Appendix \ref{appendixB} where we set $\lambda_1=\lambda_2$ and $k=2$ \footnote{Notice that there are some typos in \cite{He:2019vzf}, the exact coupling in its section 3 and section 5.1 should be $\frac{\lambda}{3}$ instead of $\lambda$.}. 

Under the conformal mapping $\eqref{equ1.11}$ for $k=2$, the stress tensors transform as 
\[
T(w)=\left(\frac{dz}{dw}\right)^2 T(z)+\frac{c}{12}\{z,w\},\quad \bar{T}(\bar{w})=\left(\frac{d\bar{z}}{d\bar{w}}\right)^2 \bar{T}(\bar{z})+\frac{c}{12}\{\bar{z},\bar{w}\}, \label{equ2.2}
\]
where 
$$\{z,w\}=\frac{z^{\prime\prime\prime}}{z^\prime}-\frac{3}{2}\frac{z^{\prime\prime 2}}{z^{\prime 2}}, \ \left(z'=\frac{dz}{dw}\right)$$ 
is the Schwarzian derivative, and $c$ is the central charge. The $T\bar{T}$ operator thus transforms as
\[
T\bar{T}(w,\bar{w})=\frac{1}{4 z ^2}\frac{1}{4 \bar{z}^2}\left( T(z)+\frac{c}{8 z^2}\right)\left( \bar{T}(\bar{z})+\frac{c}{8 \bar{z}^2}\right).\label{equ2.3}
\]
By utilizing this transformation formula and performing an expansion around $\lambda=0$, we obtain expressions for the $zero^{th}$ and first-order terms in $\lambda$:
\[
&-\left( \Delta S_{A,0}^{(2)}+\Delta S_{A,\lambda}^{(2)}\right)\nonumber\\
=&\log\left( \prod\limits_{i=1}^4\left|\frac{dw_i}{dz_i}\right|^{-2h}\frac{\la\mo(w_{1},\bar{w}_{1})\mo^{\dagger}(w_{2},\bar{w}_{2})\mo(w_{3},\bar{w}_{3})\mo^{\dagger}(w_{4},\bar{w}_{4})\ra_{\Sigma_1}}{\left(\la \mo (w_1,\bar{w}_1)\mo^{\dagger}(w_2,\bar{w}_2)\ra_{\Sigma_1}\right)^2}\right)\nonumber\\
&+2\lambda\int d^2 z\frac{1}{4\left|z\right|^2}\frac{\la\left( T(z)+\frac{c}{8 z^2}\right)\left( \bar{T}(\bar{z})+\frac{c}{8 \bar{z}^2}\right)\mo(w_{1},\bar{w}_{1})\mo^{\dagger}(w_{2},\bar{w}_{2})\mo(w_{3},\bar{w}_{3})\mo^{\dagger}(w_{4},\bar{w}_{4})\ra_{\Sigma_1}}{\la\mo(w_{1},\bar{w}_{1})\mo^{\dagger}(w_{2},\bar{w}_{2})\mo(w_{3},\bar{w}_{3})\mo^{\dagger}(w_{4},\bar{w}_{4})\ra_{\Sigma_1}}\nonumber\\
&-2\lambda\int d^2 w\frac{\la T\bar{T}(w,\bar{w}) \mo (w_1,\bar{w}_1)\mo^{\dagger}(w_2,\bar{w}_2)\ra_{\Sigma_1}}{\la  \mo (w_1,\bar{w}_1)\mo^{\dagger}(w_2,\bar{w}_2)\ra_{\Sigma_1}}+\mo(\lambda^2).\label{equ2.4}
\]
Let us focus on the large $c$ case. At the leading order, that is the $\mo(c^2)$ order, Eq. $\eqref{equ2.4}$ is evaluated as 
\[
2\lambda\frac{c^2}{64}\int d^2 z \frac{1}{4\left|z\right|^6}=6\lambda\frac{2\pi c^2}{64}\int_{0}^{\infty} d\rho \frac{1}{4\rho^5}.
\label{equ2.5}
\]
To regularize this formula, we introduce the IR cutoff and UV cutoff by replacing $(0,\infty)$ with $(\frac{1}{\Tilde{\Lambda}},\Lambda)$  and we obtain
\[
2\lambda\frac{c^2}{64}\int d^2 z \frac{1}{4\left|z\right|^6}\overset{\text{cutoff}}{\longrightarrow} 6\lambda\frac{2\pi c^2}{64}\frac{\Tilde{\Lambda}^4}{16}.
\label{equ2.6}\]
At the next order, which is of order $\mo(c)$, Eq. $\eqref{equ2.4}$ can be evaluated as
\[
&\frac{c}{8} 2\lambda\int d^2 z\frac{1}{4\left|z\right|^2}\frac{\la\left( \frac{1}{\bar{z}^2}T(z)+\frac{1}{z^2}\bar{T}(\bar{z})\right)\mo(w_{1},\bar{w}_{1})\mo^{\dagger}(w_{2},\bar{w}_{2})\mo(w_{3},\bar{w}_{3})\mo^{\dagger}(w_{4},\bar{w}_{4})\ra_{\Sigma_1}}{\la\mo(w_{1},\bar{w}_{1})\mo^{\dagger}(w_{2},\bar{w}_{2})\mo(w_{3},\bar{w}_{3})\mo^{\dagger}(w_{4},\bar{w}_{4})\ra_{\Sigma_1}}\nonumber\\
%=&2\lambda\frac{c}{32}\int d^2 z \frac{1}{\left|z\right|^2}\frac{1}{\bar{z}^2}\big(\sum\limits_{i=1}^4\frac{h}{(z-z_i)^2}-\frac{2h}{(z-z_1)(z_1-z_3)}+\frac{2h}{(z-z_3)(z_1-z_3)}\nonumber\\
%&-\frac{2h}{(z-z_2)(z_2-z_4)}+\frac{2h}{(z-z_4)(z_2-z_4)}\nonumber\\
%&+[\frac{1}{z-z_1}(-\frac{1}{z_{13}}+\frac{1}{z_{12}})+\frac{1}{z-z_2}(-\frac{1}{z_{24}}-\frac{1}{z_{12}})+\frac{1}{z-z_3}(\frac{1}{z_{34}}+\frac{1}{z_{13}})+\frac{1}{z-z_{4}}(-\frac{1}{z_{34}}+\frac{1}{z_{24}})]\frac{\eta\partial_{\eta}G(\eta,\bar{\eta})}{G(\eta,\bar{\eta})}\big)\nonumber\\
%+&2\lambda\frac{c}{32}\int d^2 z \frac{1}{\left|z\right|^2}\frac{1}{z^2}\big(\sum\limits_{i=1}^4\frac{h}{(\bar{z}-\bar{z}_i)^2}-\frac{2h}{(\bar{z}-\bar{z}_1)(\bar{z}_1-\bar{z}_3)}+\frac{2h}{(\bar{z}-\bar{z}_3)(\bar{z}_1-\bar{z}_3)}\nonumber\\
%&-\frac{2h}{(\bar{z}-{z}_2)({z}_2-\bar{z}_4)}+\frac{2h}{(\bar{z}-\bar{z}_4)(\bar{z}_2-\bar{z}_4)}\nonumber\\
%&+[\frac{1}{\bar{z}-\bar{z}_1}(-\frac{1}{\bar{z}_{13}}+\frac{1}{\bar{z}_{12}})+\frac{1}{\bar{z}-\bar{z}_2}(-\frac{1}{\bar{z}_{24}}-\frac{1}{\bar{z}_{12}})+\frac{1}{\bar{z}-z_3}(\frac{1}{\bar{z}_{34}}+\frac{1}{\bar{z}_{13}})+\frac{1}{\bar{z}-\bar{z}_{4}}(-\frac{1}{\bar{z}_{34}}+\frac{1}{\bar{z}_{24}})]\frac{\bar{\eta}\partial_{\bar{\eta}}G(\eta,\bar{\eta})}{G(\eta,\bar{\eta})}\big)\nonumber\\
=&2\lambda\frac{c\pi}{64}\left(\frac{h}{\left|z_1\right|^4}+\frac{h}{\left|z_2\right|^4}+\frac{h}{\left|z_3\right|^4}+\frac{h}{\left|z_4\right|^4}-\frac{2h z_1}{z_{13}\left|z_1\right|^4}+\frac{2h z_3}{z_{13}\left|z_3\right|^4}-\frac{2h z_2}{z_{24}\left|z_2\right|^4}+\frac{2h z_4}{z_{24}\left|z_4\right|^4}\right.\nonumber\\
&\left.+\left[\frac{z_1}{\left|z_1\right|^4}(-\frac{1}{z_{13}}+\frac{1}{z_{12}})+\frac{z_2}{\left|z_2\right|^4}(-\frac{1}{z_{24}}-\frac{1}{z_{12}})+\frac{z_3}{\left|z_3\right|^4}(\frac{1}{z_{34}}+\frac{1}{z_{13}})+\frac{z_4}{\left|z_4\right|^4}(-\frac{1}{z_{34}}+\frac{1}{z_{24}})\right]\frac{\eta\partial_{\eta}G(\eta,\bar{\eta})}{G(\eta,\bar{\eta})}\right)\nonumber\\
&+2\lambda\frac{c\pi}{64}\left(\frac{h}{\left|\bar{z}_1\right|^4}+\frac{h}{\left|\bar{z}_2\right|^4}+\frac{h}{\left|\bar{z}_3\right|^4}+\frac{h}{\left|\bar{z}_4\right|^4}-\frac{2h \bar{z}_1}{\bar{z}_{13}\left|\bar{z}_1\right|^4}+\frac{2h \bar{z}_3}{\bar{z}_{13}\left|\bar{z}_3\right|^4}-\frac{2h \bar{z}_2}{\bar{z}_{24}\left|\bar{z}_2\right|^4}+\frac{2h \bar{z}_4}{\bar{z}_{24}\left|\bar{z}_4\right|^4}\right.\nonumber\\
&\left.+\left[\frac{\bar{z}_1}{\left|\bar{z}_1\right|^4}(-\frac{1}{\bar{z}_{13}}+\frac{1}{\bar{z}_{12}})+\frac{\bar{z}_2}{\left|\bar{z}_2\right|^4}(-\frac{1}{\bar{z}_{24}}-\frac{1}{\bar{z}_{12}})+\frac{\bar{z}_3}{\left|\bar{z}_3\right|^4}(\frac{1}{\bar{z}_{34}}+\frac{1}{\bar{z}_{13}})+\frac{\bar{z}_4}{\left|\bar{z}_4\right|^4}(-\frac{1}{\bar{z}_{34}}+\frac{1}{\bar{z}_{24}})\right]\frac{\bar{\eta}\partial_{\bar{\eta}}G(\eta,\bar{\eta})}{G(\eta,\bar{\eta})}\right).
\label{equ2.7}
\]
The derivation of this equation is shown in Appendix \ref{appendixA}. When we take the late-time limit $t\rightarrow\infty$, the conformal block in the four-point correlation function is 
\[
G(\eta,\bar{\eta})\sim \mathcal{F}_{00}(1-\eta)^{-2h}\bar{\eta}^{-2h},
\label{equ2.8}
\]
where $(\mathcal{F}_{00})^{-1}=d_{\mo}$ is the quantum dimension of $\mo$ . 

Substitute \eqref{equ2.8} into the \eqref{equ2.7} and we find the correction to the second pseudo-R\'enyi entropy in large c limit at the late time:
\[
&-\Delta S_{A,\lambda}^{(2)}=6\lambda\frac{2\pi c^2}{64}\frac{\Tilde{\Lambda}^4}{16}+2\lambda\frac{4 (x_1^2-x_2^2)h c\pi}{64}\nn\\
&\left(\frac{ -t^2 \left(\sqrt{-t-x_1} \sqrt{t-x_1}+\sqrt{-t-x_2} \sqrt{t-x_2}\right)}{(-t-x_1)^{3/2} (t-x_1)^{3/2} (-t-x_2)^{3/2} (t-x_2)^{3/2} \left(\sqrt{-t-x_1}+\sqrt{-t-x_2}\right) \left(\sqrt{t-x_1}-\sqrt{t-x_2}\right)}\right.\nn\\
+&\left.\frac{ x_1^2 \sqrt{-t-x_2} \sqrt{t-x_2}+x_2^2 \sqrt{-t-x_1} \sqrt{t-x_1}}{(-t-x_1)^{3/2} (t-x_1)^{3/2} (-t-x_2)^{3/2} (t-x_2)^{3/2} \left(\sqrt{-t-x_1}+\sqrt{-t-x_2}\right) \left(\sqrt{t-x_1}-\sqrt{t-x_2}\right)}\right)\nn\\
&+\mo(\epsilon,c^0,\lambda^2).
\label{equ2.9}
%2\lambda\frac{h c\pi}{64}\frac{4 (x_1^2-x_2^2) \left(-t^2 \left(\sqrt{-t-x_1} \sqrt{t-x_1}+\sqrt{-t-x_2} \sqrt{t-x_2}\right)+x_1^2 \sqrt{-t-x_2} \sqrt{t-x_2}+x_2^2 \sqrt{-t-x_1} \sqrt{t-x_1}\right)}{(-t-x_1)^{3/2} (t-x_1)^{3/2} (-t-x_2)^{3/2} (t-x_2)^{3/2} \left(\sqrt{-t-x_1}+\sqrt{-t-x_2}\right) \left(\sqrt{t-x_1}-\sqrt{t-x_2}\right)}.
\]
%Together with the result at leading order \eqref{equ2.6}, the correction to the second pseudo-R\'enyi entropy
This result is universal in all RCFTs. However, if the two insertion coordinates coincide, Eq. \eqref{equ2.9} does not yield the same result as the second R\'enyi entropy derived in \cite{He:2019vzf}. This discrepancy arises primarily because in \cite{He:2019vzf}, we initially set $x_1\to x_2$ and perform an expansion in $\epsilon$, whereas in the case of the second pseudo-R\'enyi entropy, we first expand in $\epsilon$ and then take the limit $x_1\to x_2$. Notably, the two limits, $\epsilon\to0$ and $x_1\to x_2$, do not commute, as demonstrated in \cite{He:2023eap}.

Since the two insertion coordinates are different, the conformal block $G(\eta,\bar{\eta})$ may diverge for some models at the early time. The rest of this subsection will discuss the early-time correction to the second pseudo-R\'enyi entropy in $T\bar T$-deformed CFTs for the free scalar and Ising model as examples.

\textbf{Example 1}. The free scalar

The free scalar can be described by the operator $\mo=\frac{1}{\sqrt{2}}(e^{\frac{i}{2}\phi}+e^{-\frac{i}{2}\phi})$, whose conformal dimension is $h=\bar h=\frac{1}{8}$.
The conformal block of free scalar can be written as
\[
G(\eta,\bar{\eta})=\frac{1+\left|\eta\right|+\left|1-\eta\right|}{2\sqrt{\left|\eta\right|\left|1-\eta\right|}}. \label{equ2.10}
\]

Substituting \eqref{equ2.10} into \eqref{equ2.7}, we find that the corrections to the second pseudo-R\'enyi entropy of free scalar at the early time in large c limit is 
\[
-\Delta S_{A,\lambda}^{(2)}=&6\lambda\frac{2\pi c^2}{64}\frac{\Tilde{\Lambda}^4}{16}+\frac{1}{x_1^2 x_2^2 \left(x_1-x_2\right){}^4}\left(4 x_1^6-16 x_2 x_1^5+2 x_2^2 \left(\left(\frac{\left(x_1-x_2\right){}^4}{x_1^2 x_2^2}\right){}^{3/4}+14\right) x_1^4\right.\nn\\
+&\left(\sqrt{-x_1} \sqrt{-x_2} \left(\sqrt{\frac{x_1^2+4 \sqrt{-x_1} \sqrt{-x_2} x_1+6 x_2 x_1+x_2^2+4 \sqrt{-x_1} \sqrt{-x_2} x_2}{x_1 x_2}}\right.\right.\nn\\
-&\left.\sqrt{\frac{4 \sqrt{-x_2} \left(-x_1\right){}^{3/2}+4 \sqrt{-x_1} \left(-x_2\right){}^{3/2}+x_1^2+x_2^2+6 x_1 x_2}{x_1 x_2}}\right) \left(\frac{\left(x_1-x_2\right){}^4}{x_1^2 x_2^2}\right){}^{3/4}\nn\\
+&\left.4 x_2 \left(\left(\frac{\left(x_1-x_2\right){}^4}{x_1^2 x_2^2}\right){}^{3/4}-8\right)\right) x_2^2 x_1^3\nn\\
+&\left(\sqrt{-x_1} \sqrt{-x_2} \left(\sqrt{\frac{x_1^2+4 \sqrt{-x_1} \sqrt{-x_2} x_1+6 x_2 x_1+x_2^2+4 \sqrt{-x_1} \sqrt{-x_2} x_2}{x_1 x_2}}\right.\right.\nn\\
-&\left.\sqrt{\frac{4 \sqrt{-x_2} \left(-x_1\right){}^{3/2}+4 \sqrt{-x_1} \left(-x_2\right){}^{3/2}+x_1^2+x_2^2+6 x_1 x_2}{x_1 x_2}}\right) \left(\frac{\left(x_1-x_2\right){}^4}{x_1^2 x_2^2}\right){}^{3/4}\nn\\
+&\left.\left.2 x_2 \left(\left(\frac{\left(x_1-x_2\right){}^4}{x_1^2 x_2^2}\right){}^{3/4}+14\right)\right) x_2^3 x_1^2-16 x_2^5 x_1+4 x_2^6\right). \label{equ2.11}
%\frac{4 x_1^6-16 x_2 x_1^5+2 x_2^2 \left(\left(\frac{\left(x_1-x_2\right){}^4}{x_1^2 x_2^2}\right){}^{3/4}+14\right) x_1^4+\left(\sqrt{-x_1} \sqrt{-x_2} \left(\sqrt{\frac{x_1^2+4 \sqrt{-x_1} \sqrt{-x_2} x_1+6 x_2 x_1+x_2^2+4 \sqrt{-x_1} \sqrt{-x_2} x_2}{x_1 x_2}}-\sqrt{\frac{4 \sqrt{-x_2} \left(-x_1\right){}^{3/2}+4 \sqrt{-x_1} \left(-x_2\right){}^{3/2}+x_1^2+x_2^2+6 x_1 x_2}{x_1 x_2}}\right) \left(\frac{\left(x_1-x_2\right){}^4}{x_1^2 x_2^2}\right){}^{3/4}+4 x_2 \left(\left(\frac{\left(x_1-x_2\right){}^4}{x_1^2 x_2^2}\right){}^{3/4}-8\right)\right) x_2^2 x_1^3+\left(\sqrt{-x_1} \sqrt{-x_2} \left(\sqrt{\frac{x_1^2+4 \sqrt{-x_1} \sqrt{-x_2} x_1+6 x_2 x_1+x_2^2+4 \sqrt{-x_1} \sqrt{-x_2} x_2}{x_1 x_2}}-\sqrt{\frac{4 \sqrt{-x_2} \left(-x_1\right){}^{3/2}+4 \sqrt{-x_1} \left(-x_2\right){}^{3/2}+x_1^2+x_2^2+6 x_1 x_2}{x_1 x_2}}\right) \left(\frac{\left(x_1-x_2\right){}^4}{x_1^2 x_2^2}\right){}^{3/4}+2 x_2 \left(\left(\frac{\left(x_1-x_2\right){}^4}{x_1^2 x_2^2}\right){}^{3/4}+14\right)\right) x_2^3 x_1^2-16 x_2^5 x_1+4 x_2^6}{x_1^2 x_2^2 \left(x_1-\text{x2}\right){}^4}
\]
%where
%\[
%&u=\sqrt{\frac{x_1^2+4 (-x_1)^{3/2} \sqrt{-x_2}+6 x_1 x_2+4 \sqrt{-x_1} (-x_2)^{3/2}+x_2^2}{x_1 x_2}},\nn\\
%&v=\sqrt{\frac{x_1^2+6 x_1 x_2+4 \sqrt{-x_1} {x_1} \sqrt{-x_2}+4 \sqrt{-x_1} x_2 \sqrt{-x_2}+x_2^2}{x_1 x_2}}.
%\]
Notice that the results \eqref{equ2.9} and $\eqref{equ2.11}$ are highly different from that in \cite{He:2019vzf} because the insertion coordinates of two states generating the transition matrix are different. %However, if we take the limit $x_1\rightarrow\ x_2$ in \eqref{equ2.7}, the results above come back to the results in the second R\'enyi entropy, which is shown in \cite{He:2019vzf}.

\textbf{Example 2}. Ising model

The $2n$-point correlation function of the spin operator in the Ising model can be written as
\[
\la\sigma(z_1,\bar{z}_1)\dots\sigma(z_{2n},\bar{z}_{2n})\ra^2=\frac{1}{2^n}\sum\limits_{\substack{\epsilon_i=\pm 1 \\ i=1,\dots,2n\\ \sum\epsilon_i=0}}\prod\limits_{i<j}\left|z_i-z_j\right|^{\epsilon_i \epsilon_j /2}
\]
The first-order corrections to the second pseudo-R\'enyi entropy of the spin operator at the early time is 
\[\label{eq34}
-&\Delta S_{A,\lambda}^{(2)}=6\lambda\frac{2\pi c^2}{64}\frac{\Tilde{\Lambda}^4}{16}+2\lambda\frac{c\pi}{64}\left(
\frac{-\left(x_1^2\right){}^{3/2}+3 \sqrt{x_2^2} x_1^2+3 \sqrt{x_1^2} x_2^2+4 x_1 x_2 \left(\sqrt[4]{x_1^2}+\sqrt[4]{x_2^2}\right){}^2}{4 \left(x_1^2\right){}^{9/8} \sqrt{-2 \sqrt{-x_1} \sqrt{-x_2}-x_1-x_2} \sqrt{2 \sqrt{-x_1} \sqrt{-x_2}-x_1-x_2} \left(x_2^2\right){}^{9/8}}\right.\nn\\
+&\left.\frac{2 x_2^2 \sqrt[4]{x_1^2}+2 \left(x_1^2\right){}^{5/4}-\left(x_2^2\right){}^{5/4}}{4 \left(x_1^2\right){}^{9/8} \sqrt{-2 \sqrt{-x_1} \sqrt{-x_2}-x_1-x_2} \sqrt{2 \sqrt{-x_1} \sqrt{-x_2}-x_1-x_2} \left(x_2^2\right){}^{7/8}}+\frac{4}{x_1^2}+\frac{4}{x_2^2}
\right)+\mo(\epsilon,c^0,\lambda^2).
%&-\Delta S_{A,\lambda}^{(2)}=6\lambda\frac{2\pi c^2}{64}\frac{\Tilde{\Lambda}^4}{16}+2\lambda\frac{c\pi}{64}\left(\frac{16(x_1x_2)^{\frac{1}{4}}(x_1^2+x_2^2)}{4(x_1x_2)^{\frac{9}{4}}}+\right.\nn\\
%&\left. \frac{x_1^3+x_1^2(-7x_2+2\sqrt{x_1x_2})+x_2^2(x_2+2\sqrt{x_1x_2})+x_1x_2(-7x_2+8\sqrt{x_1x_2})}{4(x_1x_2)^{\frac{9}{4}}\left|x_1-x_2\right|}\right)+\mo(\epsilon,c^0,\lambda^2).
\]

Since all correlation functions in the Ising model are known, one can easily calculate the correction to the pseudo-R\'enyi entropy in $T\bar{T}$-deformed CFTs at the early time. As Eq. (\ref{eq34}) shows, the earlier time behavior of the pseudo-R\'enyi entropy depends on the initial setup and details of the theories. 
\section{Pseudo entropy in $T\bar{T}$-deformed CFTs }\label{section4}
In the previous section, we perform a perturbative calculation of the second pseudo-R\'enyi entropy in $T\bar{T}$-deformed CFT at the late time and corrections to the second pseudo-R\'enyi entropy of free scalar and Ising model at the early time. However, to obtain the correction to the pseudo entropy, it is reasonable to examine the $k^{\rm th}$ pseudo-R\'enyi entropy and perform an analytic continuation for $k$ goes to $1$ in $T\bar{T}$-deformed CFT. As demonstrated earlier, the conformal block for the second pseudo-R\'enyi entropy did not have a universal structure at the early stage. Therefore, in this section, firstly, we will investigate the $k^{\rm th}$ R\'enyi entropy at both the early and the late time in $T\bar{T}$-deformed CFTs. Then we only consider the $T\bar{T}$-deformation of the $k^{\rm th}$ pseudo-R\'enyi entropy at the late time.
%In the previous section, we compute the second pseudo-R\'enyi entropy in $T\bar{T}$-deformed CFT perturbatively at the late time and the corrections to the second pseudo-R\'enyi entropy of free scalar at the early time. However, to derive the pseudo-entropy, it is reasonable to consider the $k$-th pseudo-R\'enyi entropy and take $k$ analytic continuation to $1$ in $T\bar{T}$-deformed CFT. As shown above, the conformal block for pseudo-R\'enyi entropy does not have a universal form at the early time, so in this section, we firstly consider $k$-th R\'enyi entropy at the late and early time, and then we will only consider the $T\bar{T}$-deformation of $k$-th pseudo-R\'enyi entropy at the late time. 
\subsection{$k^{\rm th}$ R\'enyi entropy in $T\bar{T}$-deformed CFTs}\label{section4.1}
Under the conformal map $\eqref{equ1.11}$, the $T\bar{T}$ operators transform as
\[
T(w)=\frac{1}{k^2 z^{2k-2}}\left(T(z)+\frac{c(k^2-1)}{24}\frac{1}{z^2}\right),\quad \bar{T}(\bar{w})=\frac{1}{k^2 \bar{z}^{2k-2}}\left(\bar{T}(\bar{z})+\frac{c(k^2-1)}{24}\frac{1}{\bar{z}^2}\right).\label{equ4.1}
\]
The correction to the undeformed R\'enyi entropy seen in appendix \ref{appendixB} is
\[
\Delta S_{A,\lambda}^{(k)}=\frac{k\lambda}{k-1}\int_{\Sigma_1} d^2 w\left(\frac{\la T\bar{T}(w,\bar{w})\mo_1...\mo^\dagger_{2k}\ra_{\Sigma_k}}{\la \mo_1...\mo_{2k}^\dagger\ra_{\Sigma_k}}-\frac{\la T\bar{T}(w,\bar{w})\mo_1\mo^\dagger_{2}\ra_{\Sigma_1}}{\la \mo_1\mo_{2}^\dagger\ra_{\Sigma_1}}\right),\label{equ4.2}
\]
where  $\mo(z_i,\bar{z}_i)$ is denoted by $\mo_i$. Notice that if we take all primary operators as identity operators, \eqref{equ4.2} becomes the R\'enyi entropy for vacuum states in $T\bar{T}$-deformed CFTs, and it matches the result in \cite{Chen:2018eqk} precisely.

Using the conformal transformation \eqref{equ1.11} between $\Sigma_k$ and $\Sigma_1$, we can rewrite this formula as
\[
&\Delta S_{A,\lambda}^{(k)}=\frac{k\lambda}{k-1}\int_{\Sigma_1} d^2 z\frac{1}{k^2\left|z\right|^{2k-2}}\nn\\
&\left(\frac{\left\la (T(z)+\frac{c(k^2-1)}{24}\frac{1}{z^2})(\bar{T}(\bar{z})+\frac{c(k^2-1)}{24}\frac{1}{\bar{z}^2})\mo_1(z_1,\bar{z}_1)...\mo^\dagger_{2k}(z_{2k},\bar{z}_{2k})\right\ra_{\Sigma_1}}{\la\mo_1(z_1,\bar{z}_1)...\mo^\dagger_{2k}(z_{2k},\bar{z}_{2k})\ra_{\Sigma_1}}\right. 
- \left.\frac{\la T\bar{T}(w,\bar{w})\mo_1\mo^\dagger_{2}\ra_{\Sigma_1}}{\la \mo_1\mo_{2}^\dagger\ra_{\Sigma_1}}\right).\label{equ4.6}
\]
We still focus on the large $c$ case. At the leading order $\mo(c^2)$, by introducing the IR cutoff $\frac{1}{\Tilde{\Lambda}}$, the correction to the $k^{\rm th}$ R\'enyi entropy is
\[
\frac{k}{k-1}\lambda \frac{c^2(k^2-1)^2}{576}\int d^2 z \frac{1}{k^2\left|z\right|^{2k+2}}\overset{\text{cutoff}}{\longrightarrow}3k\lambda\frac{c^2(k^2-1)^2\pi \Tilde{\Lambda}^{2k}}{576 k^3}.\label{vacuum}
\]
The derivation of this formula is shown in  Appendix \ref{appendixA}.
At the next-to-leading order, the integral in the first line of \eqref{equ4.6} can be divided into two parts at the late time. The first part, due to $T(z)$ acting on the $2k$-point correlation function, can be written as 
\[
&\Delta S_{A,\lambda,part1}^{(k)}\equiv\frac{k\lambda}{k-1}\int_{\Sigma_1} d^2 z\frac{1}{k^2\left|z\right|^{2k-2}}\frac{c(k^2-1)}{24}\frac{1}{\bar{z}^2}\frac{\la (T(z)\mo_1(z_1,\bar{z}_1)...\mo^\dagger_{2k}(z_{2k},\bar{z}_{2k})\ra_{\Sigma_1}}{\la\mo_1(z_1,\bar{z}_1)...\mo^\dagger_{2k}(z_{2k},\bar{z}_{2k})\ra_{\Sigma_1}}\nn\\
=&\frac{k}{k-1}\lambda\frac{1}{k^2}\frac{c(k^2-1)}{24}\int d^2 z\frac{1}{\left|z\right|^{2k-2}\bar{z}^2}\frac{1}{\la\mo(z_{1},\bar{z}_{1})\mo^{\dagger}(z_{2},\bar{z}_{2})\dots\mo(z_{2k-1},\bar{z}_{2k-1})\mo^{\dagger}(z_{2k},\bar{z}_{2k})\ra_{\Sigma_1}}\nn\\
&\left(\sum\limits_{i=1}^{2k}\frac{h}{(z-z_i)^2}+\sum\limits_{j=0}^{k-2}\left(\frac{\partial_{2j+1}}{z-z_{2j+1}}+\frac{\partial_{2j+4}}{z-z_{2j+4}}\right)+\frac{\partial_{2}}{z-z_{2}}+\frac{\partial_{2k-1}}{z-z_{2k-1}}\right)\nonumber\\
&\times\la\mo(z_{1},\bar{z}_{1})\mo^{\dagger}(z_{2},\bar{z}_{2})\dots\mo(z_{2k-1},\bar{z}_{2k-1})\mo^{\dagger}(z_{2k},\bar{z}_{2k})\ra_{\Sigma_1}\nonumber\\
=&\frac{k}{k-1}\lambda \frac{1}{k^2}\frac{c(k^2-1)}{24}\bigg(\sum\limits_{i=1}^{2k}\frac{h\pi}{k}\frac{1}{\left|z_i\right|^{2k}}+\sum\limits_{j=0}^{k-2}\frac{\pi}{k}\left(\frac{-2h}{z_{2j+1}-z_{2j+4}}\frac{z_{2j+1}}{\left|z_{2j+1}\right|^{2k}}+\frac{2h}{z_{2j+1}-z_{2j+4}}\frac{z_{2j+4}}{\left|z_{2j+4}\right|^{2k}}\right)\nonumber\\
&+\frac{\pi}{k}\left(\frac{-2h}{z_2-z_{2k-1}}\frac{z_2}{\left|z_{2}\right|^{2}}+\frac{2h}{z_2-z_{2k-1}}\frac{z_{2k-1}}{\left|z_{2k-1}\right|^{2}}\right)\bigg)\label{equ4.3}
\]
where in the last equality, we use the fact 
\[
\lim\limits_{t\to\infty}(z_{2i+4}-z_{2i+1})\simeq \ \frac{w_2-w_1}{kt}e^{2\pi i\frac{i+1}{k}}t^{\frac{1}{k}}\simeq 0,
\]
 and the fact that the holomorphic part of the $2k$-point correlation function can be written as
\[
&\la\mo(z_1)\mo^\dagger(z_2)\dots\mo(z_{2k-1})\mo^{\dagger}(z_{2k})\ra_{\Sigma_1} \nn\\
\sim &(F_{00}[\mo])^{k-1}\la\mo(z_1)\mo^\dagger(z_4)\ra_{\Sigma_1}\dots\la\mo(z_{2k+1})\mo^{\dagger}(z_{2k+4})\ra_{\Sigma_1}.
\]
The second part due to $\bar{T}(\bar{z})$ acting on the $2k$-point correlation function can be written as 
\[
&\Delta S_{A,\lambda,part2}^{(k)}\equiv\frac{k\lambda}{k-1}\int_{\Sigma_1} d^2 z\frac{1}{k^2\left|z\right|^{2k-2}}\frac{c(k^2-1)}{24}\frac{1}{z^2}\frac{\la (\bar{T}(z)\mo_1(z_1,\bar{z}_1)...\mo^\dagger_{2k}(z_{2k},\bar{z}_{2k})\ra_{\Sigma_1}}{\la\mo_1(z_1,\bar{z}_1)...\mo^\dagger_{2k}(z_{2k},\bar{z}_{2k})\ra_{\Sigma_1}}\nn\\
=&\frac{k}{k-1}\lambda\frac{1}{k^2}\frac{c(k^2-1)}{24}\int d^2 z\frac{1}{\left|z\right|^{2k-2} z^2}\frac{1}{\la\mo(z_{1},\bar{z}_{1})\mo^{\dagger}(z_{2},\bar{z}_{2})\dots\mo(z_{2k-1},\bar{z}_{2k-1})\mo^{\dagger}(z_{2k},\bar{z}_{2k})\ra_{\Sigma_1}}\nn\\
&\left(\sum\limits_{i=1}^{2k}\frac{\bar{h}}{(\bar{z}-\bar{z}_i)^2}+\sum\limits_{j=0}^{k-1}\left(\frac{\bar{\partial}_{2j+1}}{\bar{z}-\bar{z}_{2j+1}}+\frac{\bar{\partial}_{2j+2}}{\bar{z}-\bar{z}_{2j+2}}\right)\right)\nonumber\\
&\times\la\mo(w_{1},\bar{w}_{1})\mo^{\dagger}(w_{2},\bar{w}_{2})\dots\mo(w_{2k-1},\bar{w}_{2k-1})\mo^{\dagger}(w_{2k},\bar{w}_{2k})\ra_{\Sigma_1}\nonumber\\
=&\frac{k}{k-1}\lambda\frac{1}{k^2}\frac{c(k^2-1)}{24}\left(\sum\limits_{i=1}^{2k}\frac{\bar{h}\pi}{k}\frac{1}{\left|\bar{z}_i\right|^{2k}}+\sum\limits_{j=0}^{k-1}\frac{\pi}{k}\left(\frac{-2\bar{h}}{\bar{z}_{2j+1}-\bar{z}_{2j+2}}\frac{\bar{z}_{2j+1}}{\left|\bar{z}_{2j+1}\right|^{2k}}+\frac{2\bar{h}}{\bar{z}_{2j+1}-\bar{z}_{2j+2}}\frac{\bar{z}_{2j+2}}{\left|\bar{z}_{2j+2}\right|^{2k}}\right)\right), \label{equ4.4}
\]
where in the last equality, we use the fact 
\[
\lim\limits_{t\to\infty}(\bar{z}_{2i+2}-\bar{z}_{2i+1})\simeq \ \frac{w_1-w_2}{kt}e^{-2\pi i\frac{i+\frac{1}{2}}{k}}t^{\frac{1}{k}}\simeq 0,
\]
and the fact that the anti-holomorphic part of the $2k$-point correlation function can be written as
\[
&\la\mo(\bar{z}_1)\mo^\dagger(\bar{z}_2)\dots\mo(\bar{z}_{2k-1})\mo^{\dagger}(\bar{z}_{2k})\ra_{\Sigma_1} \nn\\
\sim &(F_{00}[\mo])^{k-1}\la\mo(\bar{z}_1)\mo^\dagger(\bar{z}_2)\ra_{\Sigma_1}\dots\la\mo(\bar{z}_{2k-1})\mo^{\dagger}(\bar{z}_{2k})\ra_{\Sigma_1}.
\]
Now we replace all coordinates in \eqref{equ4.3} and \eqref{equ4.4} with
\[
z_{2k+1}=&\text{e}^{2\pi i\frac{k+1/2}{n}}(-x-t+i\e)^{\frac{1}{n}},\quad \bar z_{2k+1}=\text{e}^{-2\pi i\frac{k+1/2}{n}}(-x+t-i\e)^{\frac{1}{n}},\nn\\
z_{2k+2}=&\text{e}^{2\pi i\frac{k+1/2}{n}}(-x-t-i\e)^{\frac{1}{n}},\quad \bar z_{2k+2}=\text{e}^{-2\pi i\frac{k+1/2}{n}}(-x+t+i\e)^{\frac{1}{n}},\quad (k=0,...,n-1).
\]
We get the correction to the $k^{\rm th}$ R\'enyi entropy in the $T\bar{T}$-deformed CFT at the late time
\[
\Delta S^{(k)}_{A,\lambda}=3k\lambda\frac{c^2(k^2-1)^2\pi \Tilde{\Lambda}^{2k}}{576 k^3}-\frac{k}{k-1}\lambda\frac{h\pi c(k^2-1)}{24k^3}\frac{4k^2 t}{(t-x)(t+x)^2}+\mo(\epsilon,c^0,\lambda^2).\label{equ4.7}
\]
We next consider the integral \eqref{equ4.6} at the early time. The coordinates and the $2k$-point correlation function have the following property:
\[
&\lim\limits_{t\to 0 }(z_{2i+2}-z_{2i+1})\simeq 0, \quad\lim\limits_{t\to 0}(\bar{z}_{2i+2}-\bar{z}_{2i+1})\simeq 0, \nn\\
&\la\mo(z_1,\bar{z}_1)\mo^\dagger(z_2,\bar{z}_2)\dots\mo(z_{2k-1},\bar{z}_{2k-1})\mo^{\dagger}(z_{2k},\bar{z}_{2k})\ra_{\Sigma_1} \nn\\
\sim &\la\mo(z_1,\bar{z}_1)\mo^\dagger(z_2,\bar{z}_2)\ra_{\Sigma_1}\dots\la\mo(z_{2k-1},\bar{z}_{2k-1})\mo^{\dagger}(z_{2k},\bar{z}_{2k})\ra_{\Sigma_1}.\label{earlyproperty}
%\la\mo(\bar{z}_1)\mo^\dagger(\bar{z}_2)\dots\mo(\bar{z}_{2k-1})\mo^{\dagger}(\bar{z}_{2k})\ra_{\Sigma_1} 
%\sim \la\mo(\bar{z}_1)\mo^\dagger(\bar{z}_2\ra_{\Sigma_1}\dots\la\mo(\bar{z}_{2k-1})\mo^{\dagger}(\bar{z}_{2k})\ra_{\Sigma_1}
\]
We can get the correction to the $k^{\rm th}$ R\'enyi entropy at the early time.
\[
\Delta S_{A,\lambda}^{(k)}=3k\lambda\frac{c^2(k^2-1)^2\pi \Tilde{\Lambda}^{2k}}{576 k^3}-\frac{k}{k-1}\lambda\frac{h\pi c(k^2-1)}{24k^3}\frac{8k^2 t^2}{(t^2-x^2)^2}+\mo(\epsilon,c^0,\lambda^2). \label{equ4.8}
\]
Results $\eqref{equ4.7}$ and $\eqref{equ4.8}$ are precisely the results we find in \cite{He:2019vzf} if we take the interval of $A$ in \cite{He:2019vzf} to infinity and set $k=2$.
Taking the limit of $k\to1$ for \eqref{equ4.7} and \eqref{equ4.8}, we obtain the early-time and late-time behavior of the entanglement entropy under the $T\bar{T}$-deformation up to the first-order, respectively. Namely,
\begin{align}\Delta S_{A,\lambda}=
\begin{cases}
-\lambda\frac{2\pi hc}{3}\frac{t^2}{\left(t^2-x^2\right)^2},& t<|x|,\\
-\lambda\frac{\pi hc}{3}\frac{t}{\left(t-x\right)\left(t+x\right)^2},& t>|x|.
\end{cases}\label{eecorrection}
\end{align}
In this scenario, the $T\bar{T}$ deformation does not alter the late-time dynamics of $\Delta S_{A,\lambda}$, which exhibits a temporal evolution following a $\frac{1}{t^2}$ pattern. It is important to note that a singularity emerges in \eqref{eecorrection} at $t=x$, and this singularity arises from the derivative acting on the $2k$-point correlation function, which discontinuously changes at that point.
\subsection{$k^{\rm th}$ pseudo-R\'enyi entropy in $T\bar{T}$-deformed CFTs}
Thanks to the property of the $2k$-point correlation function \eqref{earlyproperty} at the early time, we can derive the correction to the $k^{\rm th}$ R\'enyi entropy in $T\bar{T}$-deformed CFTs. However, this property is no longer valid for the pseudo entropy since the two insertion coordinates $x_1$ and $x_2$ are different. Therefore we only consider the $T\bar{T}$-deformation of the $k^{\rm th}$ pseudo-R\'enyi entropy at the late time in this subsection.

Combining \eqref{equ4.3} and \eqref{equ4.4}, and replacing all the coordinates with \eqref{A=INFzcoordin}, we get the correction to the $k^{\rm th}$ pseudo-R\'enyi entropy in the $T\bar{T}$-deformed CFT,
\[
&\Delta S_{A,\lambda}^{(k)}=3k\lambda\frac{c^2(k^2-1)^2\pi \Tilde{\Lambda}^{2k}}{576 k^3}\nn\\
+&\frac{k}{k-1}\lambda\frac{h\pi c(k^2-1)}{24k^3}\left(\frac{2k\left((-t+x_1)^{\frac{1}{k}}(t-x_1)^{\frac{1}{k}}-e^{\frac{2\pi i}{k}}(-t-x_2)^{\frac{1}{k}}(t-x_2)^{\frac{1}{k}}\right)(x_1^2-x_2^2)}{(t^2-x_1^2)\left((-t-x_1)^{\frac{1}{k}}-e^{\frac{2\pi i}{k}}(-t-x_2)^{\frac{1}{k}}\right)\left((t-x_1)^{\frac{1}{k}}-(t-x_2)^{\frac{1}{k}}\right)(t^2-x_2^2)}\right)\nn\\
+&\mo(\epsilon,c^0,\lambda^2).\label{equ4.5}
\]
The result \eqref{equ4.5} precisely matches the result we find in the previous section \ref{section3} if we take $k=2$. 

Taking the limit of $k\to 1$ for \eqref{equ4.5}, we obtain the late-time correction of the pseudo entropy under the $T\bar{T}$-deformation up to the first-order which is
\[
\Delta S_{A,\lambda}=\frac{\lambda c h \pi}{6}\frac{((-t-x_1)(t-x_1)-(-t-x_2)(t-x_2))(x_1^2-x_2^2)}{(t^2-x_1^2)(-x_1^2+x_2^2)(t^2-x_2^2)}.
\]
In the limit of late time, it is evident that $\Delta S_{A,\lambda}\sim \frac{1}{ t^2}$. Therefore, the first-order $T\bar{T}$ deformation does not alter the late-time behavior of the pseudo-entropy, which is consistent with the late-time behavior of the excess entanglement entropy derived in \eqref{eecorrection}. Moreover, there exist singularities at $t=x_1$ or $t=x_2$, which emerge due to the non-smooth behavior of the correlation function at those specific points. These singularities exhibit similarities to the finding discussed in the previous subsection.
\section{Pseudo entropy in $T\bar{T}$-deformed CFTs with different deformed states}\label{section5}
In the previous section, we have explicitly shown the correction of the $k^{\rm th}$ pseudo-R\'enyi entropy after the $T\bar{T}$-deformation, up to the first order of the deformation coefficient. Recent research about the $T\bar{T}$-deformed CFTs has focused on how to glue two different AdS spacetimes \cite{Kawamoto:2023wzj, Apolo:2023vnm} and its holographic dual. Inspired by the cAdS/dCFT correspondence, in this section, we will consider the pseudo-R\'enyi entropy for states with different $T\bar{T}$-deformation on the CFT side. 

\begin{figure}[h]
\centering
\includegraphics[width=0.7\linewidth]{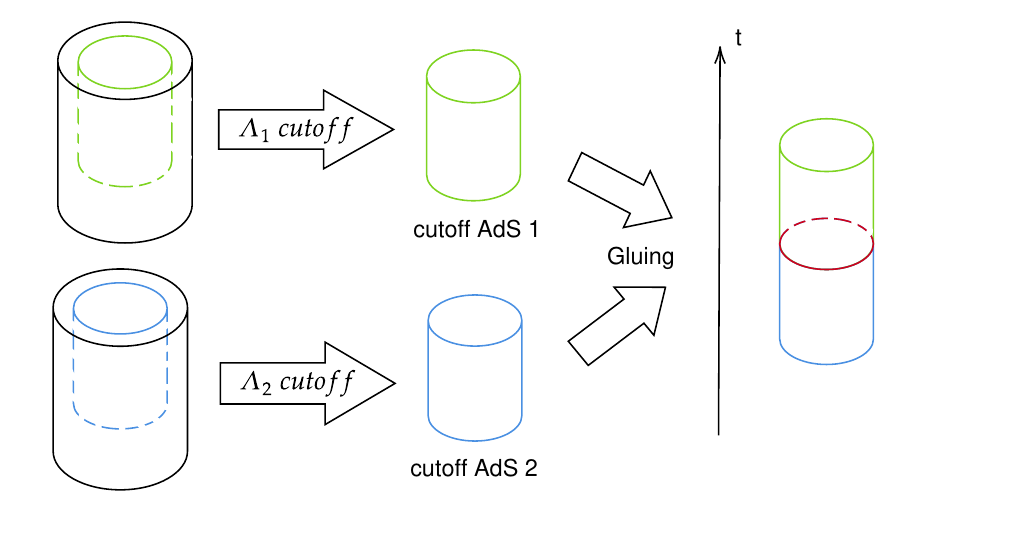}
\caption{Gluing of two cutoff AdS spacetimes along their space-like boundaries}\label{fig1}
\end{figure}

We first consider the transition matrix for the vacuum state taking the form
\[
\mathcal{T}^{1|2}\equiv\frac{|\Omega_{\lambda_1}\rangle\langle\Omega_{\lambda_2}|}{\langle\Omega_{\lambda_2}|\Omega_{\lambda_1}\rangle},
\]
where the vacuum states are deformed by $T\bar{T}$ operators with different coupling constants $\lambda_1$ and $\lambda_2$, and this transition matrix is defined in the boundary theory of a glued AdS spacetime with two cutoff AdS spacetimes gluing together along a space-like hypersurface, as shown in Figure \ref{fig1}.
 In the rest of this section, we mainly consider the case that the deformation parameters, $\lambda_1$ and $\lambda_2$, are infinitesimally small. For $\lambda_1\sim 0$ and $\lambda_2 \sim 0$,
 %, the spectrum of the deformed CFT has a one-to-one correspondence towards the spectrum of the undeformed theory. 
 the deformed vacuum states, $|\Omega_{\lambda_1}\ra$ and $|\Omega_{\lambda_2}\ra$, can be expanded according to the complete basis of the undeformed Hilbert space $\mathcal{H}$. Therefore, in the sense of perturbation theory, the deformation of the $k^{\rm th}$ pseudo-R\'enyi entropy can be derived through the replica method shown in appendix \ref{appendixB}, which is
\[
\Delta S_{A,\lambda_1,\lambda_2}^{(k)}=\frac{k}{k-1}&\left[\lambda_1\int_{\text{LHP}}d^2 w\left(\la T\bar{T}(w,\bar{w})\ra_{\Sigma_k}-\la T\bar{T}(w,\bar{w})\ra_{\Sigma_1}\right)\right. \nn\\
+&\left.\lambda_2\int_{\text{UHP}}d^2 w\left(\la T\bar{T}(w,\bar{w})\ra_{\Sigma_k}-\la T\bar{T}(w,\bar{w})\ra_{\Sigma_1}\right)\right]. \label{equ5.1}
\]
In Eq. \eqref{equ5.1}, LHP denotes the lower half-plane of $\Sigma_1$, while UHP denotes the upper half-plane of $\Sigma_1$. 

Using the transformation of the $T\bar{T}$ operator between $\Sigma_k$ and $\Sigma_1$, the integral becomes
\[
\Delta S_{A,\lambda_1,\lambda_2}^{(k)}=&\frac{k}{k-1}\left[\lambda_1\int_{\LHP}d^2 z\frac{1}{k^2 \left|z\right|^{2k-2}}\left\la \left(T(z)+\frac{c(k^2-1)}{24 z^2}\right)\left(\bar{T}(\bar{z})+\frac{c(k^2-1)}{24\bar{z}^2}\right)\right\ra_{\Sigma_1}\right.\nn\\
&\left.+\lambda_2\int_{\UHP}d^2 z \frac{1}{k^2 \left|z\right|^{2k-2}}\left\la \left(T(z)+\frac{c(k^2-1)}{24 z^2}\right)\left(\bar{T}(\bar{z})+\frac{c(k^2-1)}{24\bar{z}^2}\right)\right\ra_{\Sigma_1}\right]\nn\\
=&\frac{k}{k-1}\Big[\lambda_1\int_{\LHP} d^2 z  \frac{1}{k^2 \left|z\right|^{2k-2}}\frac{c(k^2-1)}{24 z^2}\frac{c(k^2-1)}{24\bar{z}^2}\nn\\
&+\lambda_2\int_{\UHP} d^2 z  \frac{1}{k^2 \left|z\right|^{2k-2}}\frac{c(k^2-1)}{24 z^2}\frac{c(k^2-1)}{24\bar{z}^2}\Big]\nn\\
=&\lambda_1\frac{c^2(k-1)(k+1)^2}{1152k^2}\Tilde{\Lambda}_1^{2k}+\lambda_2\frac{c^2(k-1)(k+1)^2}{1152k^2}\Tilde{\Lambda}_2^{2k},\label{equ5.2}
\]
where $\tilde\Lambda_1$ and $\tilde\Lambda_2$ are two different IR cutoffs. Since we are dealing with two distinct deformation couplings, there is no requirement for the two cutoffs to be identical. However, in the case of vacuum states, if we exchange the indices $(1\leftrightarrow 2)$, the path integral constructed through the replica method should remain invariant. This symmetry is also satisfied by the result in Eq. \eqref{equ5.2}. In the scenario where the two deformation parameters are equal, we must set $\Tilde{\Lambda}_1=\Tilde{\Lambda}_2$. Under this condition, Eq. \eqref{equ5.2} reverts to Eq. \eqref{vacuum}.

%Since we are now considering two different deformation couplings, there is no reason to require the two cutoffs to be the same. However, for the vacuum states, if we exchange the two indices $(1\leftrightarrow 2)$, the path integral building from the replica method should be invariant, and this symmetry also satisfied in the result $\eqref{equ5.2}$. If the two deformation parameters are the same, we must set $\Tilde{\Lambda}_1=\Tilde{\Lambda}_2$, and under this circumstance, \eqref{equ5.2} returns back to \eqref{vacuum}.
%T(w)=\frac{1}{k^2 z^{2k-2}}\left(T(z)+\frac{c(k^2-1)}{24}\frac{1}{z^2}\right),\quad \bar{T}(\bar{w})=\frac{1}{k^2 \bar{z}^{2k-2}}\left(\bar{T}(\bar{z})+\frac{c(k^2-1)}{24}\frac{1}{\bar{z}^2}\right)
%We now consider a transition matrix with two primary operators, where they act on different deformed vacuum states, and we set them at the same insertion coordinates, 

%|\psi_1\ra\equiv \mo(x)|\Omega_{\lambda_1}\ra,\quad |\psi_2\ra\equiv \mo(x)|\Omega_{\lambda_2}\ra
In the subsequent part of this section, we shall examine the pseudo-R\'enyi entropy derived from primary operators acting on distinct vacuum states with two different infinitesimally small constants in $T\bar{T}$-deformed CFT. The states used to construct the transition matrix are now expressed as follows:
%In the rest of this section, we will consider the pseudo-R\'enyi entropy constructed from primary operators with different deformation constants in $T\bar{T}$- deformed CFTs. States constructing the transition matrix are now taking the form:
\[
|\psi_1\rangle=\mo(w_1)|\Omega_{\lambda_1}\rangle,\quad
|\psi_2\rangle= \mo(w_2)|\Omega_{\lambda_2}\rangle.
\]
According to appendix \ref{appendixB}, the  deformed $k^{\rm th}$ pseudo-R\'enyi entropy is now taking the form
\[
\Delta S_{A,\lambda_1,\lambda_2}^{(k)}=&\frac{k}{k-1}\lambda_1\int_{\LHP~\text{of}~\Sigma_1} d^2 w\left(\frac{\la T\bar{T}(w,\bar{w})\mo_1...\mo^\dagger_{2k}\ra_{\Sigma_k}}{\la \mo_1...\mo_{2k}^\dagger\ra_{\Sigma_k}}-\frac{\la T\bar{T}(w,\bar{w})\mo_1\mo^\dagger_{2}\ra_{\Sigma_1}}{\la \mo_1\mo_{2}^\dagger\ra_{\Sigma_1}}\right)\nn\\
&+\frac{k}{k-1}\lambda_2\int_{\UHP~\text{of}~\Sigma_1} d^2 w\left(\frac{\la T\bar{T}(w,\bar{w})\mo_1...\mo^\dagger_{2k}\ra_{\Sigma_k}}{\la \mo_1...\mo_{2k}^\dagger\ra_{\Sigma_k}}-\frac{\la T\bar{T}(w,\bar{w})\mo_1\mo^\dagger_{2}\ra_{\Sigma_1}}{\la \mo_1\mo_{2}^\dagger\ra_{\Sigma_1}}\right).\label{equ5.3}
\]
Through the calculation in appendix \ref{appendixC}, we have the  correction to the $k^{\rm th}$ pseudo-R\'enyi entropy with different deformed parameters of $O(c)$ under $T\bar T$ deformation at the late time
\[
&\Delta S_{A,\lambda_1,\lambda_2}^{(k)}=\nn\\
&\frac{c(k+1)}{24 k^2}\lambda_1\Bigg\{\frac{\left((-t+x_1)^{\frac{1}{k}}(t-x_1)^{\frac{1}{k}}-e^{\frac{2\pi i}{k}}(-t-x_2)^{\frac{1}{k}}(t-x_2)^{\frac{1}{k}}\right)(x_1^2-x_2^2)}{(t^2-x_1^2)\left((-t-x_1)^{\frac{1}{k}}-e^{\frac{2\pi i}{k}}(-t-x_2)^{\frac{1}{k}}\right)\left((t-x_1)^{\frac{1}{k}}-(t-x_2)^{\frac{1}{k}}\right)(t^2-x_2^2)}\nn\\
+&\frac{1}{2} \pi \Bigg [\frac{\left(x_1-x_2\right) \left(2 t-x_1-x_2\right) \left(\left(t-x_1\right){}^{1/k}+\left(t-x_2\right){}^{1/k}\right)}{\left(t-x_1\right){}^2 \left(t-x_2\right){}^2 \left(\left(t-x_1\right){}^{1/k}-\left(t-x_2\right){}^{1/k}\right)}+\frac{2 \left(\left(-t-x_1\right){}^{\frac{1}{k}-2}-e^{\frac{2 i \pi }{k}} \left(-t-x_2\right){}^{\frac{1}{k}-2}\right)}{-\left(-t-x_1\right){}^{1/k}+e^{\frac{2 i \pi }{k}} \left(-t-x_2\right){}^{1/k}}\nn\\
+&2 t^2+2 t \left(x_1+x_2\right)  +2 \left(\frac{1}{t+x_2}+\frac{1}{t+x_1}\right)+x_1^2+x_2^2\Bigg ]\Bigg\}
+\mo(\epsilon,c^0,\lambda_1^2)\nn\\
+&\frac{c(k+1)}{24 k^2}\lambda_2\Bigg \{\frac{\left((-t+x_1)^{\frac{1}{k}}(t-x_1)^{\frac{1}{k}}-e^{\frac{2\pi i}{k}}(-t-x_2)^{\frac{1}{k}}(t-x_2)^{\frac{1}{k}}\right)(x_1^2-x_2^2)}{(t^2-x_1^2)\left((-t-x_1)^{\frac{1}{k}}-e^{\frac{2\pi i}{k}}(-t-x_2)^{\frac{1}{k}}\right)\left((t-x_1)^{\frac{1}{k}}-(t-x_2)^{\frac{1}{k}}\right)(t^2-x_2^2)}\nn\\
-&\frac{1}{2} \pi \Bigg [\frac{\left(x_1-x_2\right) \left(2 t-x_1-x_2\right) \left(\left(t-x_1\right){}^{1/k}+\left(t-x_2\right){}^{1/k}\right)}{\left(t-x_1\right){}^2 \left(t-x_2\right){}^2 \left(\left(t-x_1\right){}^{1/k}-\left(t-x_2\right){}^{1/k}\right)}+\frac{2 \left(\left(-t-x_1\right){}^{\frac{1}{k}-2}-e^{\frac{2 i \pi }{k}} \left(-t-x_2\right){}^{\frac{1}{k}-2}\right)}{-\left(-t-x_1\right){}^{1/k}+e^{\frac{2 i \pi }{k}} \left(-t-x_2\right){}^{1/k}}\nn\\
+&2 t^2+2 t \left(x_1+x_2\right)  +2 \left(\frac{1}{t+x_2}+\frac{1}{t+x_1}\right)+x_1^2+x_2^2\Bigg ]\Bigg\}+\mo(\epsilon,c^0,\lambda_2^2).\label{equ5.4}
\]
Notice that there are also some possible cutoff terms with $\epsilon_1$ and $\epsilon_2$ corresponding to the LHP and UHP separately and we have regularized them through minimal subtraction seen in \eqref{appendix3.4}\text{-}\eqref{appendix3.11}. When the two deformation parameters are identical, we require that both cutoffs, $\epsilon_1$ and $\epsilon_2$, are set to the same value $\epsilon$. In this case, formula \eqref{equ5.4}, including the ignored  cutoff terms, will precisely give rise to the result in Eq. \eqref{equ4.5}. However, when $\lambda_1\neq \lambda_2$, the correction to the $k^{\rm th}$ pseudo-R\'enyi entropy may involve additional nontrivial terms compared to $\eqref{equ4.5}$. 

\begin{figure}[h]
\centering
\includegraphics[width =0.9\linewidth]{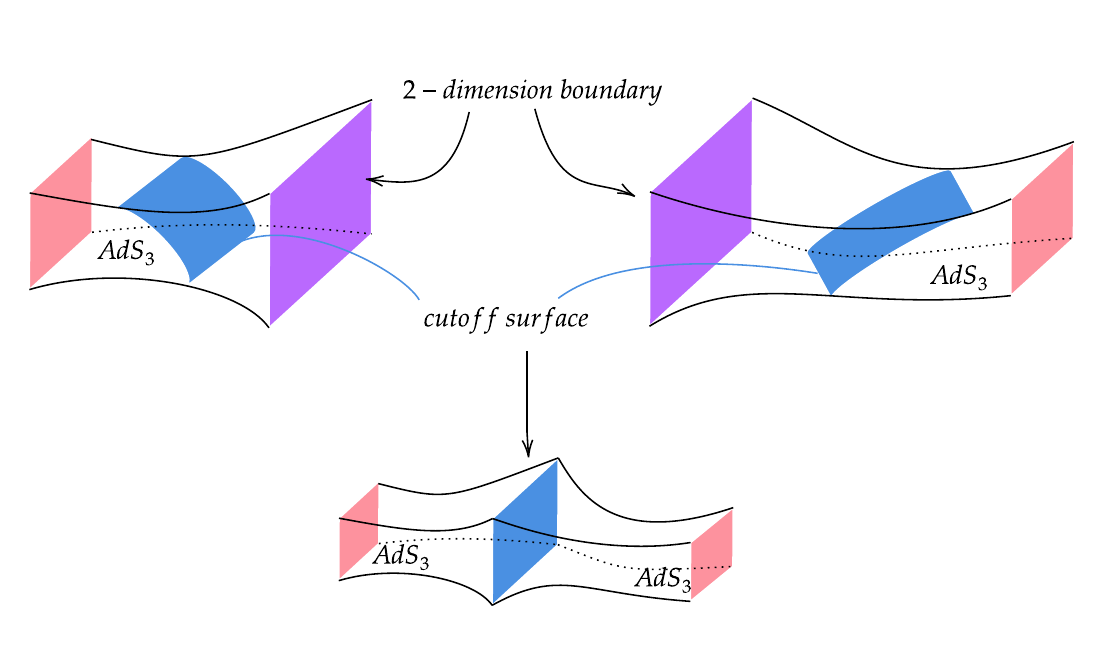}
\caption{Gluing of two cutoff $\text{AdS}_3$ along their time-like boundaries}\label{fig2}
\end{figure}

When the deformation parameters are large enough, the two deformed theories denoted as CFT$^1$ and CFT$^2$ are far from the undeformed theory, so states in CFT$^1$ and CFT$^2$ belong to two different Hilbert space $\mathcal{H}_1$ and $\mathcal{H}_2$. The gluing procedure in figure \ref{fig1} gives rise to an invalid transition matrix, since ordinarily the transition matrix is defined with states in the same Hilbert space \cite{Nakata:2020luh} \footnote{We thank the anonymous referee for pointing this out to us.}. In order to establish a well-defined transition matrix, we could consider the tensor product theory CFT$^1\otimes$CFT$^2$ which has been extensively investigated in \cite{Ferko:2022dpg}.
 From the holographic view, such tensor product theory may relate to the gluing of two cutoff spacetimes along their time-like cutoff boundaries (or branes) as shown in figure \ref{fig2}. 
In the rest of this section, we will concentrate on pseudo entropy constructed from states of the tensor product CFT$^1\otimes$CFT$^2$.

Let us first consider a sufficiently simple tensor product theory as a warm-up. The Hamiltonian of the theory is given by
\[
H=H_1\otimes I_2+I_1\otimes H_2\label{hamiltonian},
\]
where $H_i$ (i=1,2) is the Hamiltonian of the corresponding deformed theory CFT$^i$ and $I_i$ (i=1,2) is the identity operator for CFT$^i$. The inner product of $|\varphi\ra=|\varphi_1\ra\otimes|\varphi_2\ra$ is defined as 
\[
\la\varphi|\varphi\ra=\la\varphi_1|\varphi_1\ra\otimes\la\varphi_2|\varphi_2\ra=\la\varphi_1|\varphi_1\ra \cdot\la\varphi_2|\varphi_2\ra.
\]
We then construct two states of the tensor product theory by acting tensor product operators on the deformed vacuum state $|\Omega_{\lambda_1}\rangle\otimes|\Omega_{\lambda_2}\rangle$
\[
|\psi_1\ra=\mo(w_1)\otimes\mo'(w_2) |\Omega_{\lambda_1}\ra\otimes|\Omega_{\lambda_2}\ra=\mo(w_1)|\Omega_{\lambda_1}\ra\otimes\mo'(w_2) |\Omega_{\lambda_2}\ra,\nn\\
|\psi_2\ra=\mo(w_2)\otimes\mo'(w_1) |\Omega_{\lambda_1}\ra\otimes|\Omega_{\lambda_2}\ra=\mo(w_2) |\Omega_{\lambda_1}\ra\otimes\mo'(w_1)|\Omega_{\lambda_2}\ra.
\]
Before calculating the time evolution of pseudo-R\'enyi entropy of $|\psi_1\ra$ and $|\psi_2\ra$ using the replica method, we would like to show specifically the path integral in the tensor product theory. As an explicit example, we first consider the correlation function of $\mo\otimes\mo'$ in the tensor product theory, which is
\[
\la\mo(w_1)\otimes\mo'(w_2)\ra_{\text{CFT}^{1}\otimes\text{CFT}^2}=\frac{1}{Z}&\int\mathcal{D}\phi_1\mathcal{D}\phi_2 \mo(\phi_1)\otimes\mo'(\phi_2) \exp\{-\mathcal{L}_1-\mathcal{L}_2\}\nn\\
=&\frac{1}{Z_1}\int\mathcal{D}\phi_1\mo(\phi_1)\exp\{-\mathcal{L}_1\}\cdot\frac{1}{Z_2}\int\mathcal{D}\phi_2\mo'(\phi_2)\exp\{-\mathcal{L}_2\}\nn\\
=&\la\mo(w_1)\ra_{\text{CFT}^1}\cdot\la\mo'(w_2)\ra_{\text{CFT}^2}.
\]
The transition matrix of $|\psi_1\ra$ and $|\psi_2\ra$ is
\[
\mathcal{T}^{\psi_1 | \psi_2}=&\frac{\mo(w_1)\otimes\mo'(w_2) |\Omega_{\lambda_1}\ra\otimes|\Omega_{\lambda_2}\ra\la\Omega_1|\otimes\la\Omega_2|\mo^\dagger(w_2)\otimes{\mo'}^{\dagger}(w_1)}{\la\Omega_1|\mo^\dagger(w_2)\mo(w_1)|\Omega_1\ra\cdot\la\Omega_2|{\mo'}^\dagger(w_1)\mo'(w_2)|\Omega_2\ra}\nn\\
=&\frac{\mo(w_1) |\Omega_{\lambda_1}\ra\la\Omega_1|\mo^\dagger(w_2)}{\la\Omega_1|\mo^\dagger(w_2)\mo(w_1)|\Omega_1\ra}\otimes\frac{\mo'(w_2) |\Omega_{\lambda_2}\ra\la\Omega_2|{\mo'}^\dagger(w_1)}{\la\Omega_2|{\mo'}^\dagger(w_1)\mo'(w_2)|\Omega_2\ra},\label{trans1}
\]
and after using replica method $(\mathcal{T}_A^{\psi_1 | \psi_2})^k$ becomes
\[
(\mathcal{T}_A^{\psi_1 | \psi_2})^k=\left(\tr_{\bar A}\frac{\mo(w_1) |\Omega_{\lambda_1}\ra\la\Omega_1|\mo^\dagger(w_2)}{\la\Omega_1|\mo^\dagger(w_2)\mo(w_1)|\Omega_1\ra}\right)^k\otimes\left(\tr_{\bar A}\frac{\mo'(w_2) |\Omega_{\lambda_2}\ra\la\Omega_2|{\mo'}^\dagger(w_1)}{\la\Omega_2|{\mo'}^\dagger(w_1)\mo'(w_2)|\Omega_2\ra}\right)^k.
\]
Therefore, the correction to the $k^{\rm th}$ pseudo-R\'enyi entropy for tensor product theory can be written as 
\[
\Delta S^{(k)}_{A,\lambda_1,\lambda_2}=\frac{k}{k-1}(\lambda_1+\lambda_2)\int_{\Sigma_1} d^2 w &\left(\frac{\la T\bar{T}(w,\bar{w})\mo_1\otimes\mo'_1...\mo^\dagger_{2k}\otimes\mo'^\dagger_{2k}\ra_{\Sigma_k}}{\la \mo_1\otimes\mo'_1...\mo^\dagger_{2k}\otimes\mo'^\dagger_{2k}\ra_{\Sigma_k}}\right.\nn\\
-&\left.\frac{\la T\bar{T}(w,\bar{w})\mo_1\otimes\mo'_1\mo^\dagger_{2}\otimes\mo'^\dagger_{2}\ra_{\Sigma_1}}{\la \mo_1\otimes\mo'_1\mo^\dagger_{2}\otimes\mo'^\dagger_{2}\ra_{\Sigma_1}}\right).\label{correction of ttbar}
%+&\frac{k}{k-1}\lambda_2\int_{\Sigma_1} d^2 w\left(\frac{\la T\bar{T}(w,\bar{w})\mo_1'...{\mo'}^\dagger_{2k}\ra_{\Sigma_k}}{\la \mo_1'...{\mo'}_{2k}^\dagger\ra_{\Sigma_k}}-\frac{\la T\bar{T}(w,\bar{w})\mo'_1{\mo'}^\dagger_{2}\ra_{\Sigma_1}}{\la \mo_1'{\mo'}_{2}^\dagger\ra_{\Sigma_1}}\right)
\]
%which we have explicitly shown this quantity in the previous section.

%We conclude this section with some comments about the relation between the gluing procedure and the tensor product theory. 
In the above discussion, our focus is on the condition that the two deformed theories remain independent. However, as demonstrated in \cite{Kawamoto:2023wzj}, if we consider the gluing of two AdS spacetimes, we should add an interaction term $H_{int}$ to the Hamiltonian of tensor product theory \eqref{hamiltonian} which probably can be understood in terms of an irrelevant deformation that couples the two deformed CFTs together, where the author in \cite{Ferko:2022dpg} call it sequential flow, and this term arises due to the exchange of gravitons between two AdS spacetimes. 
%Hence, it becomes intriguing to explore the relationship between the gluing procedure and wormhole physics since exchanging gravitons between two AdS spacetimes seems to creat a channel between the two spacetimes. 
Under such flow, the Lagrangian of the tensor product theory with interactions up to the first order of deformation parameters can be expressed as \cite{Ferko:2022dpg}
\[
L=L_1+L_2+\lambda_3[(T_1+T_2)(\bar T_1+\bar T_2)]+\lambda_1 T_1\bar T_1+\lambda_2 T_2\bar T_2.\label{sfL}
\]
Due to the interaction term, the transition matrix of interest cannot be decomposed as \eqref{trans1} in this case
\[
\mathcal{T}^{\psi_1|\psi_2}=\frac{\mo(w_1)\otimes\mo'(w_2)|\Omega\rangle_{SF} ~_{SF}\langle\Omega|\mo^\dagger(w_2)\otimes\mo'^\dagger(w_1)}{_{SF}\langle\Omega|\mo^{\dagger}(w_2)\otimes\mo'^\dagger(w_1)\mo(w_1)\otimes\mo'(w_2)|\Omega\rangle_{SF}},\label{transi2}
\]
where $|\Omega\rangle_{SF}$ denotes the vacuum of the sequential-flowed theory \eqref{sfL}.
We still can expand the correlation function under sequential flow perturbatively, and the result is 
\[
\la\mo\otimes\mo'\ra_{SF}=&\frac{1}{\sqrt{Z_{SF}}}\int\mathcal{D}\phi_1\mathcal{D}\phi_2 \mo\otimes\mo' \exp\{-\mathcal{L}_1-\mathcal{L}_2-\lambda_3[(T_1+T_2)(\bar T_1+\bar T_2)]-\lambda_1 T_1\bar T_1-\lambda_2 T_2\bar T_2\}\nn\\
=&\frac{1}{\sqrt{Z_0}}\int\mathcal{D}\phi_1\mathcal{D}\phi_2 \mo\otimes\mo' \exp\{-\mathcal{L}_1-\mathcal{L}_2\}(1-\lambda_3[(T_1+T_2)(\bar T_1+\bar T_2)]-\lambda_1 T_1\bar T_1-\lambda_2 T_2\bar T_2)\nn\\
=&\la\mo\otimes\mo'\ra-\lambda_3\int \dd^2 z\la(T_1(z)+T_2(z))(\bar T_1(\bar z)+\bar T_2(\bar z))\mo\otimes\mo'\ra\nn\\
&-\lambda_1\int \dd^2 z\la T_1(z)\bar T_1(
\bar z)\mo\otimes\mo'\ra-\lambda_2\int \dd^2 z\la T_2(z)\bar T_2(
\bar z)\mo\otimes\mo'\ra
\]
where $Z_{SF}$ is the partition function under sequential flow which can be expressed as
\[
Z_{SF}=&\int\mathcal{D}\phi_1\mathcal{D}\phi_2\exp\{-\mathcal{L}_1-\mathcal{L}_2-\lambda_3[(T_1+T_2)(\bar T_1+\bar T_2)]-\lambda_1 T_1\bar T_1-\lambda_2 T_2\bar T_2\}\nn\\
=&Z_0-(\lambda_3+\lambda_1)\int \dd^2 z\la T_1(z)\bar T_1(z)\ra-(\lambda_3+\lambda_2)\int \dd^2 z\la T_2(z)\bar T_2(z)\ra\nn\\
&-\lambda_3\int \dd^2 z(\la T_1(z)\ra\la \bar T_2(\bar z)\ra+\la \bar T_1(\bar z)\ra\la  T_2(z)\ra).
\]
Using the method in section \ref{section4} we can derive the correction to the $k^{\rm th}$ pseudo-R\'enyi entropy for tensor product theory under sequential flow 
\[
&\Delta S^{(k)}_{A,\lambda_1,\lambda_2,\lambda_3}\nn\\
=&\frac{k}{k-1}(\lambda_1+\lambda_3)\int_{\Sigma_1} d^2 w\left(\frac{\la T_1\bar{T}_1(w,\bar{w})\mo_1\otimes\mo'_1...\mo^\dagger_{2k}\otimes\mo'^\dagger_{2k}\ra_{\Sigma_k}}{\la \mo_1\otimes\mo'_1...\mo^\dagger_{2k}\otimes\mo'^\dagger_{2k}\ra_{\Sigma_k}}-\frac{\la T_1\bar{T}_1(w,\bar{w})\mo_1\otimes\mo'_1\mo^\dagger_{2}\otimes\mo'^\dagger_{2}\ra_{\Sigma_1}}{\la \mo_1\otimes\mo'_1\mo^\dagger_{2}\otimes\mo'^\dagger_{2}\ra_{\Sigma_1}}\right)\nn\\
+&\frac{k}{k-1}(\lambda_2+\lambda_3)\int_{\Sigma_1} d^2 w\left(\frac{\la T_2\bar{T_2}(w,\bar{w})\mo_1\otimes\mo'_1...\mo^\dagger_{2k}\otimes\mo'^\dagger_{2k}\ra_{\Sigma_k}}{\la \mo_1\otimes\mo'_1...\mo^\dagger_{2k}\otimes\mo'^\dagger_{2k}\ra_{\Sigma_k}}-\frac{\la T_2\bar{T}_2(w,\bar{w})\mo_1\otimes\mo'_1\mo^\dagger_{2}\otimes\mo'^\dagger_{2}\ra_{\Sigma_1}}{\la \mo_1\otimes\mo'_1\mo^\dagger_{2}\otimes\mo'^\dagger_{2}\ra_{\Sigma_1}}\right)\nn\\
+&\frac{k}{k-1}\lambda_3\int_{\Sigma_1} d^2 w \left(\frac{\la T_1\bar{T_2}(w,\bar{w})\mo_1\otimes\mo'_1...\mo^\dagger_{2k}\otimes\mo'^\dagger_{2k}\ra_{\Sigma_k}}{\la \mo_1\otimes\mo'_1...\mo^\dagger_{2k}\otimes\mo'^\dagger_{2k}\ra_{\Sigma_k}}-\frac{\la T_1\bar{T}_2(w,\bar{w})\mo_1\otimes\mo'_1\mo^\dagger_{2}\otimes\mo'^\dagger_{2}\ra_{\Sigma_1}}{\la \mo_1\otimes\mo'_1\mo^\dagger_{2}\otimes\mo'^\dagger_{2}\ra_{\Sigma_1}}\right)\nn\\
+&\frac{k}{k-1}\lambda_3\int_{\Sigma_1} d^2 w \left(\frac{\la \bar T_1 T_2(w,\bar{w})\mo_1\otimes\mo'_1...\mo^\dagger_{2k}\otimes\mo'^\dagger_{2k}\ra_{\Sigma_k}}{\la \mo_1\otimes\mo'_1...\mo^\dagger_{2k}\otimes\mo'^\dagger_{2k}\ra_{\Sigma_k}}-\frac{\la \bar T_1 T_2(w,\bar{w})\mo_1\otimes\mo'_1\mo^\dagger_{2}\otimes\mo'^\dagger_{2}\ra_{\Sigma_1}}{\la \mo_1\otimes\mo'_1\mo^\dagger_{2}\otimes\mo'^\dagger_{2}\ra_{\Sigma_1}}\right),\label{correction of sq}
\]
where the correlation function containing the term such as $T_1 \bar T_2$ can be expressed as
\[
\la  T_1 \bar T_2(w,\bar{w})\mo_1\otimes\mo'_1...\mo^\dagger_{2k}\otimes\mo{'}^\dagger_{2k}\ra=\la T_1(w)\mo_1...\mo^\dagger_{2k}\ra\cdot\la\bar T_2(\bar w)\mo'_1...\mo{'}^\dagger_{2k}\ra,
\]
and the term such as $T_1\bar T_1$ can be expressed as
\[
\la  T_1 \bar T_1(w,\bar{w})\mo_1\otimes\mo'_1...\mo^\dagger_{2k}\otimes\mo{'}^\dagger_{2k}\ra=\la T_1 \bar T_1(w,\bar{w})\mo_1...\mo^\dagger_{2k}\ra\cdot\la\bar \mo'_1...\mo{'}^\dagger_{2k}\ra.
\]
All terms in \eqref{correction of sq} can be computed similarly using the results in \eqref{vacuum},  \eqref{equ4.3} and \eqref{equ4.4}. However, comparing \eqref{correction of sq} and \eqref{correction of ttbar}, we find the result in \eqref{correction of sq} has additional corrections shown in the last two lines of \eqref{correction of sq}, which can not be derived through the renaming of $\lambda_1$ and $\lambda_2$ in \eqref{correction of ttbar}, and these corrections are totally due to the interaction between the two CFTs of the tensor product theory.
%We conclude this section with some comments. %about the relation between the gluing procedure and the tensor product theory.

%As argued in \cite{Kawamoto:2023wzj}, gluing two AdS spacetimes along the brane would create an AdS bulk spacetime without boundaries, so in order to gain insight into the connection between the gluing procedure and the sequential flow, it is essential to delve into the relationship between the boundary-less AdS spacetime and its corresponding field theory, and consider the $T\bar T$-deformation of the field theory. 
The above computations are  an initial attempt in RCFTs under the sequential flow \cite{Ferko:2022dpg}. In order to gain insight into the connection between the gluing procedure and the sequential flow, we expect to find the holographic pseudo-R\'enyi entropy of transition matrix \eqref{transi2} in the concept of ``holography without boundaries" initiated in \cite{Kawamoto:2023wzj}. The holographic calculations in glued AdS and the field-theoretic calculations in holographic CFTs under sequential flow are necessary and  will be served as our future work.

%We could further extend our research to holographic CFT, where we expect that the extended investigation in holographic CFT may offer valuable insights, particularly in relation to the boundary conditions during the gluing process \cite{Kawamoto:2023wzj, Apolo:2023vnm}. %Research  can provide a deeper understanding of the dynamics and properties of the pseudo-R\'enyi entropy in holographic setups, and potentially lead to novel approaches and considerations when dealing with gluing procedures.
%Notice that even if the $k^{\rm th}$ pseudo-R\'enyi entropy for different vacuum states acquires non-dynamical contribution in the $T\bar{T}$-deformation, the contribution still depends on the two deformation parameters, and this may give insight into gluing on two different cAdS geometries. We believe that the $k^{\rm th}$ pseudo-R\'enyi entropy for the primary operators with different deformation constants may have some nontrivial corrections depending on the two deformation coupling constants. 
 
\section{Pseudo-R\'enyi entropy in $J\bar{T}$-deformed CFTs}\label{section6}
So far, we have discussed the $T\bar{T}$-deformation of the pseudo-R\'enyi entropy. In this section, we study the $J\bar{T}$-deformation of the pseudo entropy, with $J\bar{T}$ operator constructed from a chiral $U(1)$ current $J$ and stress tensor $\bar{T}$.

The $J\bar{T}$-deformed action is the trajectory on the space of field theory satisfying
\[
\frac{d S}{d\lambda}=\int d^2 z(J\bar{T})_\lambda.
\]
Similar to the $T\bar{T}$-deformation, we regard the deformation as a perturbative theory, in which case the action can be written as
\[
S(\lambda)=S(\lambda=0)+\lambda\int d^2 z \sqrt{g} J\bar{T}+\mo(\lambda^2),
\]
where  $(J\bar{T})_{\lambda=0}$ is denoted by $J\bar{T}$ later on. The first-order correction to the correlation function by using the Ward identity is
\[
&\la\mo_{1}(z_1,\bar{z}_1)\mo_{2}(z_2,\bar{z}_2)\dots\mo_{k}(z_k,\bar{z}_k)\ra_{\lambda}\nonumber\\
 =&\lambda \int d^2 z\left\langle J \bar{T}(z, \bar{z}) \mo_1\left(z_1, \bar{z}_1\right) \mo_2\left(z_2, \bar{z}_2\right) \cdots \mo_k\left(z_k, \bar{z}_k\right)\right\rangle\\
=&\lambda\int d^2z\left(\sum\limits^{k}_{i=1}\frac{q_i}{z-z_i}\right)\left(\sum\limits^{k}_{i=1}\left(\frac{\bar{h}_i}{(\bar{z}-\bar{z}_i)^2}+\frac{\partial_{\bar{z}_i}}{\bar{z}-\bar{z}_i}\right)\right)\nonumber\\
&\times \la\mo_{1}(z_1,\bar{z}_1)\mo_{2}(z_2,\bar{z}_2)\dots\mo_{k}(z_k,\bar{z}_k)\ra.
\]
We then calculate the correction to the $k^{\rm th}$ pseudo-R\'enyi entropy in the $J\bar{T}$-deformed CFT. 

Under the conformal map \eqref{equ1.11}, the $U(1)$ current and the stress tensor transform as
\[
J\bar{T}(w,\bar{w})=\frac{dz}{dw}J(z)\frac{1}{4\bar{z}^2}\left(\bar{T}(\bar{z})+\frac{c}{8\bar{z}^2}\right).
\]
The correction to the undeformed $k^{\rm th}$ pseudo-R\'enyi entropy can be derived similarly from that in the previous section. Thus we have
\[
\Delta S_{A,\lambda}^{(k)}=\frac{k\lambda}{k-1}\int_{\Sigma_1} d^2 w\left(\frac{\la J\bar{T}(w,\bar{w})\mo_1...\mo^\dagger_{2k}\ra_{\Sigma_k}}{\la \mo_1...\mo_{2k}^\dagger\ra_{\Sigma_k}}-\frac{\la J\bar{T}(w,\bar{w})\mo_1\mo^\dagger_{2}\ra_{\Sigma_1}}{\la \mo_1\mo_{2}^\dagger\ra_{\Sigma_1}}\right).
\]
We still focus on the large $c$ case. At $\mo(c)$, $\Delta S_{A,\lambda}^{(k)}$ at the late time can be written as
\[
\Delta S_{A,\lambda}^{(k)}=&\frac{k\lambda}{k-1}\int d^2 z \frac{c}{8\bar{z}^2}\frac{1}{k \bar{z}^{k-1}}\frac{\la J(z)\mo_{1}(z_1,\bar{z}_1)\mo^{\dagger}_{2}(z_2,\bar{z}_2)\dots\mo_{2k-1}(z_{2k-1},\bar{z}_{2k-1})\mo^{\dagger}_{2k}(z_{2k},\bar{z}_{2k})\ra_{\Sigma_1}}{\la\mo_{1}(z_1,\bar{z}_1)\mo^{\dagger}_{2}(z_2,\bar{z}_2)\dots\mo_{2k-1}(z_{2k-1},\bar{z}_{2k-1})\mo^{\dagger}_{2k}(z_{2k},\bar{z}_{2k})\ra_{\Sigma_1}}\nonumber\\
=&\frac{k\lambda}{k-1}\int d^2 z \frac{c}{8\bar{z}^2}\frac{1}{k \bar{z}^{k-1}}\left(\sum\limits^{2k}_{i=1}\frac{q_i}{z-z_i}\right)\nonumber\\
=&\frac{2\lambda \pi c}{8(k-1)}\sum\limits_{i=1}^{2k}\int_{\left|z_i\right|}^{\infty}\frac{q_i z_i^k}{\rho^{2k+1}}=\frac{2\lambda \pi c}{16k(1-k)}\sum\limits_{i=1}^{2k}\frac{q_i z_i^k}{\left|z_i\right|^{2k}}.\label{equ6.1}
\]
Substituting \eqref{A=INFzcoordin} into \eqref{equ6.1}, we get the first-order corrections to the $k^{\rm th}$ pseudo-R\'enyi entropy in $J\bar{T}$-deformed CFT,
\[
\Delta S_{A,\lambda}^{(k)}=\frac{2\lambda \pi c}{8(k-1)}\sum\limits_{i=0}^{k-1}(\frac{q_{2i+1}}{-t+x_1}+\frac{q_{2i+2}}{-t+x_2})+\mo(c^0,\lambda^2).
\]
Specifically, we consider the $k=2$ case,
\[
\Delta S_{A,\lambda}^{(2)}
=2\lambda\frac{c\pi}{32}\left(\frac{q_1+q_3}{-t+x_1}+\frac{q_2+q_4}{-t+x_2}\right)+\mo(c^0,\lambda^2).\label{equ6.2}
\]
If we take the limit $x_1\rightarrow x_2$, \eqref{equ6.2} returns to the result we found in \cite{He:2019vzf}.  Using the method mentioned in subsection \ref{section4.1}, one can also calculate the late-time and early-time corrections to the entanglement entropy in $J\bar{T}$-deformed CFTs.

\section{Conclusion and prospect}\label{conclusion}
This paper employs perturbation theory to investigate the pseudo-R\'enyi entropy associated with locally excited states in the two-dimensional $T\bar{T}/J\bar{T}$ deformed CFTs. Our primary objective is to derive nontrivial corrections to the pseudo-R\'enyi entropy of the undeformed theory. Initially, we provide a comprehensive review of the calculation of two-point and four-point correlation functions using the Ward identity in deformed CFTs and the pseudo-R\'enyi entropy obtained through the replica method in undeformed CFTs. We then derive the leading order corrections to the second pseudo-R\'enyi entropy and find that it may depend on the introduced infrared (IR) cutoff and may also depend on the ultraviolet (UV) cutoff, depending on whether the subsystem $A$ is finite. Furthermore, at order $\mathcal{O}(c)$, we discover that the second pseudo-R\'enyi entropy may acquire nontrivial corrections in the deformed CFTs at the late time, as demonstrated in Eqs. \eqref{equ2.9} and \eqref{equ6.2}. However, the early-time correction becomes more intricate, and we illustrate its behavior using the free scalar and Ising models as specific examples.

In addition, we extend our investigation to the $k^{\rm th}$ R\'enyi entropy, examining its behavior both at the early and late times as expressed in Eqs. \eqref{equ4.7} and \eqref{equ4.8}, respectively. Subsequently, we calculate the $k$-th pseudo-R\'enyi entropy of locally excited states at the late time, using the fact that the 2$k$-point correlation function can be deformed as the product of $k$ two-point functions. We observe nontrivial corrections to the $k^{\rm th}$ pseudo-R\'enyi entropy at the late time, as presented in Eq. \eqref{equ4.5}. Notably, this result aligns precisely with the findings from section \ref{section3}.

Finally, drawing inspiration from gluing two AdS spacetimes with distinct cutoffs and the correspondence between cutoff AdS (cAdS) and $T\bar T$-deformed CFT (dCFT), we separately investigate the pseudo-R\'enyi entropy for vacuum states characterized by distinct $T\bar{T}$-deformation parameters, and for primary states acting on different deformed vacuum states. At the leading order, we find that the corrections to the pseudo-R\'enyi entropy may depend on two different cutoffs, as indicated in Eq. \eqref{equ5.2}. For the pseudo-R\'enyi entropy associated with two  primary operators acting on different deformed vacuum states, we obtain additional nontrivial terms in Eq. \eqref{equ5.4} compared to Eq. \eqref{equ4.5}. 

Our primary focus centers on the computation of corrections induced by $T\bar T/J\bar T$-deformation to the pseudo-R\'enyi entropy in RCFTs. Specifically, by employing the $T\bar{T}/J\bar{T}$ deformation, it is possible to directly compute corrections to the pseudo-R\'enyi entropy for different local operators in various theoretical frameworks, including Liouville CFT, holographic CFT, and other relevant approaches. Moreover, exploring the definition of an operator within the framework of $T\bar{T}$-deformation, constructing pseudo-R\'enyi entropy for locally excited states in the AdS context, connecting distinct cutoff AdS geometries, and calculating the pseudo-R\'enyi entropy for locally excited states with different deformation parameters on the cAdS side could provide valuable insights. However, these captivating avenues remain unexplored and await future research endeavors.

\section*{Acknowledgements}
We thank Yang Liu, Hongfei Shu, Yuan Sun, Hongan Zeng, and Long Zhao for the valuable discussions. S.H. would appreciate the financial support from the Fundamental Research Funds for the Central Universities, the Max Planck Partner group, and the Natural Science Foundation of China Grants (No.~12075101, No.~12235016). J.Y. is partially supported by the Natural Science Foundation of China Grants (No. 11971322).

\appendix
\section{Details of integrals in $T\bar T$-deformed pseudo-R\'enyi entropy }\label{appendixA}
In this appendix, we evaluate the integral of \eqref{equ2.7} in detail. It mainly takes the following forms
\[
\int d^2 z\frac{1}{\left|z\right|^2}\frac{1}{z^2}\frac{1}{(\bar{z}-\bar{z}_i)}\quad{\rm and }\quad \int d^2 z\frac{1}{\left|z\right|^2}\frac{1}{z^2}\frac{1}{(\bar{z}-\bar{z}_i)^2}\label{a.1}
\]
as well as  their complex conjugate. Now we evaluate them accordingly. Firstly
\[
&\int d^2 z\frac{1}{\left|z\right|^2}\frac{1}{z^2}\frac{1}{(\bar{z}-\bar{z}_i)^2}
=\int_{0}^{\infty}d\rho\int_0^{2\pi} d \theta \frac{1}{\rho^3 e^{2i\theta}}\frac{1}{(\rho e^{-i\theta}-\bar{z}_i)^2}.
%=&\int d\rho\frac{\frac{i\rho}{e^{i \theta}\bar{z_i}-\rho}+i\log(\bar{z}_i-\rho e^{-i\theta})}{\rho^5}\Big|_{\theta=0}^{\theta=2\pi}\nonumber\\
%=&\int_{\left|\bar{z}_i\right|}^{\infty}\frac{2\pi}{\rho^5}=\frac{2\pi}{4\left|\bar{z}_i\right|^4}
\]
Integrating out $\theta$, we find
\[
&\int d^2 z\frac{1}{\left|z\right|^2}\frac{1}{z^2}\frac{1}{(\bar{z}-\bar{z}_i)^2}=\int d\rho\frac{\frac{i\rho}{e^{i \theta}\bar{z_i}-\rho}+i\log(\bar{z}_i-\rho e^{-i\theta})}{\rho^5}\Big|_{\theta=0}^{\theta=2\pi}.
\]
Let us consider the process where $\theta$ runs from $0$ to $2\pi$. When $\left|\frac{\rho}{\bar{z}_i}\right|>1$, $\rho e^{-i\theta}$ circles counterclockwise around $\bar{z}_i$. Hence $\log(\bar{z}_i-\rho e^{-i\theta})$ contributes a factor $-2\pi i$. Therefore we have 
\[
\log(\bar{z}_i-\rho e^{-i\theta})=\begin{cases}
0\ \ &|\frac{\rho}{\bar{z}_i}|<1 \\
-2\pi i\ \ &|\frac{\rho}{\bar{z}_i}|>1
\end{cases}.
\]
The integral becomes
\[
&\int d^2 z\frac{1}{\left|z\right|^2}\frac{1}{z^2}\frac{1}{(\bar{z}-\bar{z}_i)^2}
=\int_{\left|\bar{z}_i\right|}^{\infty}\frac{2\pi}{\rho^5}d\rho=\frac{2\pi}{4\left|\bar{z}_i\right|^4}.
\]
Secondly the other integral in \eqref{a.1} can be evaluated as
\[
&\int d^2 z\frac{1}{\left|z\right|^2}\frac{1}{z^2}\frac{1}{(\bar{z}-\bar{z}_i)}
=\int_{0}^{\infty}d\rho\int_0^{2\pi} d \theta \frac{1}{\rho^3 e^{2i\theta}}\frac{1}{(\rho e^{-i\theta}-\bar{z}_i)}\nonumber\\
=&\int d\rho\frac{ie^{-i\theta}\rho+i\bar{z}_i\log(\bar{z}_i-\rho e^{-i\theta})}{\rho^5}\Big|_{\theta=0}^{\theta=2\pi}\int_{\left|\bar{z}_i\right|}^{\infty}\frac{2\pi\bar{z}_i}{\rho^5}d\rho=\frac{2\pi\bar{z}_i}{4\left|\bar{z}_i\right|^4}.
\]
Other integrals in \eqref{equ2.7} can be evaluated in a similar way. We find that those terms coupled with $\bar{T}$ become
\[
&\frac{c}{8} 2\lambda\int d^2 z\frac{1}{4\left|z\right|^2}\frac{\la\left( \frac{1}{z^2}\bar{T}(\bar{z})\right)\mo(w_{1},\bar{w}_{1})\mo^{\dagger}(w_{2},\bar{w}_{2})\mo(w_{3},\bar{w}_{3})\mo^{\dagger}(w_{4},\bar{w}_{4})\ra_{\Sigma_1}}{\la\mo(w_{1},\bar{w}_{1})\mo^{\dagger}(w_{2},\bar{w}_{2})\mo(w_{3},\bar{w}_{3})\mo^{\dagger}(w_{4},\bar{w}_{4})\ra_{\Sigma_1}}\nonumber\\
=&2\lambda\frac{c}{32}\int d^2 z \frac{1}{\left|z\right|^2}\frac{1}{z^2}\left(\sum\limits_{i=1}^4\frac{h}{(\bar{z}-\bar{z}_i)^2}-\frac{2h}{(\bar{z}-\bar{z}_1)(\bar{z}_1-\bar{z}_3)}+\frac{2h}{(\bar{z}-\bar{z}_3)(\bar{z}_1-\bar{z}_3)}\right. \nonumber\\
&-\frac{2h}{(\bar{z}-{z}_2)({z}_2-\bar{z}_4)}+\frac{2h}{(\bar{z}-\bar{z}_4)(\bar{z}_2-\bar{z}_4)}\nonumber\\
&\left.+[\frac{1}{\bar{z}-\bar{z}_1}(-\frac{1}{\bar{z}_{13}}+\frac{1}{\bar{z}_{12}})+\frac{1}{\bar{z}-\bar{z}_2}(-\frac{1}{\bar{z}_{24}}-\frac{1}{\bar{z}_{12}})+\frac{1}{\bar{z}-z_3}(\frac{1}{\bar{z}_{34}}+\frac{1}{\bar{z}_{13}})+\frac{1}{\bar{z}-\bar{z}_{4}}(-\frac{1}{\bar{z}_{34}}+\frac{1}{\bar{z}_{24}})]\frac{\bar{\eta}\partial_{\bar{\eta}}G(\eta,\bar{\eta})}{G(\eta,\bar{\eta})}\right)\nonumber\\
=&2\lambda\frac{c\pi}{64}\left(\frac{h}{\left|\bar{z}_1\right|^4}+\frac{h}{\left|\bar{z}_2\right|^4}+\frac{h}{\left|\bar{z}_3\right|^4}+\frac{h}{\left|\bar{z}_4\right|^4}-\frac{2h \bar{z}_1}{\bar{z}_{13}\left|\bar{z}_1\right|^4}+\frac{2h \bar{z}_3}{\bar{z}_{13}\left|\bar{z}_3\right|^4}-\frac{2h \bar{z}_2}{\bar{z}_{24}\left|\bar{z}_2\right|^4}+\frac{2h \bar{z}_4}{\bar{z}_{24}\left|\bar{z}_4\right|^4} \right.\nonumber\\
&\left.+[\frac{\bar{z}_1}{\left|\bar{z}_1\right|^4}(-\frac{1}{\bar{z}_{13}}+\frac{1}{\bar{z}_{12}})+\frac{\bar{z}_2}{\left|\bar{z}_2\right|^4}(-\frac{1}{\bar{z}_{24}}-\frac{1}{\bar{z}_{12}})+\frac{\bar{z}_3}{\left|\bar{z}_3\right|^4}(\frac{1}{\bar{z}_{34}}+\frac{1}{\bar{z}_{13}})+\frac{\bar{z}_4}{\left|\bar{z}_4\right|^4}(-\frac{1}{\bar{z}_{34}}+\frac{1}{\bar{z}_{24}})]\frac{\bar{\eta}\partial_{\bar{\eta}}G(\eta,\bar{\eta})}{G(\eta,\bar{\eta})}\right).
\]
Those terms coupled with $T$ are 
\[
&\frac{c}{8} 2\lambda\int d^2 z\frac{1}{4\left|z\right|^2}\frac{\la\left( \frac{1}{\bar{z}^2}T(z)\right)\mo(w_{1},\bar{w}_{1})\mo^{\dagger}(w_{2},\bar{w}_{2})\mo(w_{3},\bar{w}_{3})\mo^{\dagger}(w_{4},\bar{w}_{4})\ra_{\Sigma_1}}{\la\mo(w_{1},\bar{w}_{1})\mo^{\dagger}(w_{2},\bar{w}_{2})\mo(w_{3},\bar{w}_{3})\mo^{\dagger}(w_{4},\bar{w}_{4})\ra_{\Sigma_1}}\nonumber\\
=&2\lambda\frac{c}{32}\int d^2 z \frac{1}{\left|z\right|^2}\frac{1}{\bar{z}^2}\left(\sum\limits_{i=1}^4\frac{h}{(z-z_i)^2}-\frac{2h}{(z-z_1)(z_1-z_3)}+\frac{2h}{(z-z_3)(z_1-z_3)}\right. \nonumber\\
&-\frac{2h}{(z-z_2)(z_2-z_4)}+\frac{2h}{(z-z_4)(z_2-z_4)}\nonumber\\
&\left.+[\frac{1}{z-z_1}(-\frac{1}{z_{13}}+\frac{1}{z_{12}})+\frac{1}{z-z_2}(-\frac{1}{z_{24}}-\frac{1}{z_{12}})+\frac{1}{z-z_3}(\frac{1}{z_{34}}+\frac{1}{z_{13}})+\frac{1}{z-z_{4}}(-\frac{1}{z_{34}}+\frac{1}{z_{24}})]\frac{\eta\partial_{\eta}G(\eta,\bar{\eta})}{G(\eta,\bar{\eta})}\right)\nonumber\\
=&2\lambda\frac{c\pi}{64}\left(\frac{h}{\left|z_1\right|^4}+\frac{h}{\left|z_2\right|^4}+\frac{h}{\left|z_3\right|^4}+\frac{h}{\left|z_4\right|^4}-\frac{2h z_1}{z_{13}\left|z_1\right|^4}+\frac{2h z_3}{z_{13}\left|z_3\right|^4}-\frac{2h z_2}{z_{24}\left|z_2\right|^4}+\frac{2h z_4}{z_{24}\left|z_4\right|^4}\right.\nonumber\\
&\left.+[\frac{z_1}{\left|z_1\right|^4}(-\frac{1}{z_{13}}+\frac{1}{z_{12}})+\frac{z_2}{\left|z_2\right|^4}(-\frac{1}{z_{24}}-\frac{1}{z_{12}})+\frac{z_3}{\left|z_3\right|^4}(\frac{1}{z_{34}}+\frac{1}{z_{13}})+\frac{z_4}{\left|z_4\right|^4}(-\frac{1}{z_{34}}+\frac{1}{z_{24}})]\frac{\eta\partial_{\eta}G(\eta,\bar{\eta})}{G(\eta,\bar{\eta})}\right).
\]
We then list the results of those integrals in \eqref{equ4.3} and \eqref{equ4.4} using the same method mentioned above.
\[
\int d^2 z\frac{1}{\left|z\right|^{2k-2} \bar{z}^2}\frac{1}{(z-z_i)^2}=\frac{\pi}{k\left|z_i\right|^{2k}}.\nn\\
\int d^2 z\frac{1}{\left|z\right|^{2k-2} \bar{z}^2}\frac{1}{(z-z_i)}=\frac{\pi z_i}{k\left|z_i\right|^{2k}}.\nn\\
\int d^2 z\frac{1}{\left|\bar{z}\right|^{2k-2} z^2}\frac{1}{(\bar{z}-\bar{z}_i)^2}=\frac{\pi}{k\left|\bar{z}_i\right|^{2k}}.\nn\\
\int d^2 z\frac{1}{\left|\bar{z}\right|^{2k-2} z^2}\frac{1}{(\bar{z}-\bar{z}_i)}=\frac{\pi\bar{z}_i}{k\left|\bar{z}_i\right|^{2k}}.
\]
\section{The $T\bar{T}$-deformation of the $k^{\rm th}$ pseudo-R\'enyi entropy by the replica method}\label{appendixB}
This appendix provides a derivation of the leading-order corrections of the  $k^{\text{th}}$ pseudo-R\'enyi entropy in 2D CFTs under the  $T\bar{T}$-deformation, using the replica trick. 

We first generate the states of interest by applying a local operator to the vacuum of the $T\bar{T}$-deformed theory with varying parameters, that is
\[
|\psi_i\rangle=\mo(x_i,\tau_i)|\Omega_{\lambda_i}\ra,~(i=1,2)
\]
with $\mo(x_i,\tau_i)\equiv\mo(w_i,\bar{w}_i)=e^{H_{\lambda_i}\tau_i}\mo(x_i,\tau=0)e^{-H_{\lambda_i}\tau_i}.$ $|\Omega_{\lambda_i}\rangle$ and $H_{\lambda_i}$ denote the vacuum and the Hamiltonian of the deformed theory with the deformation parameter $\lambda_i$, respectively. By the standard replica trick, we can express the essential part of the $k^{\rm th}$ pseudo-R\'enyi entropy, namely the trace of the $k^{\rm th}$ power of the reduced transition matrix, as the following Euclidean path integral.
\[
\text{tr}[(\mathcal{T}^{\psi_1|\psi_2}_A)^k]=\left(\imineq{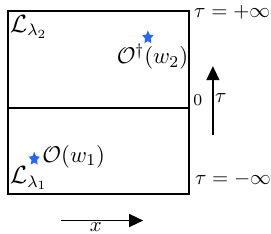}{20}\right)^{-k}\times \imineq{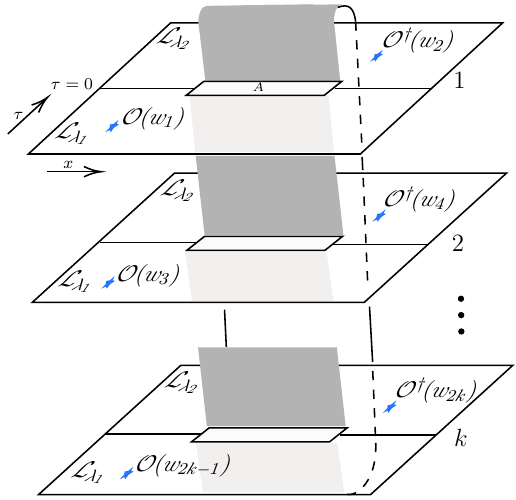}{30}.\label{trakpic}
\]
The first factor of the right-hand side of  Eq. \eqref{trakpic} corresponds to a two-point function on the complex plane $\Sigma_1$, while the second factor corresponds to a $2k$-point function on the $k$-sheeted Riemann surface $\Sigma_k$. Notably, the $T\bar{T}$-deformed theory with parameter $\lambda_1$ lives on the lower half-plane (LHP) of each sheeted surface, whereas the deformed theory with parameter $\lambda_2$ resides on the upper half-plane (UHP) of each sheeted surface.

Rewriting the above result as the correlation functions and 
expanding it according to the order of $\lambda_{1(2)}$, we obtain\footnote{We denote $(w,\bar w)$ by $(w)$ in $\mo$ and $T\bar{T}$ for simplicity.}
\[
&\tr[(\mathcal{T}_A^{\psi_1|\psi_2})^k]\nn\\
%=&\frac{\int[\dd\phi]\mo(w_1)\mo^\dagger(w_2)...\mo(w_{2k-1})\mo^\dagger(w_{2k})\exp\big\{-k\int_{\text{LHP}}\mathcal{L}_{\lambda_1}(\phi)-k\int_{\text{UHP}}\mathcal{L}_{\lambda_2}(\phi)\big\}}{\left(\int[\dd\phi]\mo(w_1)\mo^\dagger(w_2)\exp\big\{-\int_{\text{LHP}}\mathcal{L}_{\lambda_1}(\phi)-\int_{\text{UHP}}\mathcal{L}_{\lambda_2}(\phi)\big\}\right)^k}\nn\\
%=&\frac{\int[\dd\phi]\mo(w_1)\mo^\dagger(w_2)...\mo(w_{2k-1})\mo^\dagger(w_{2k})e^{-\int_{\Sigma_k}\mathcal{L}_0[\phi]}\big(1-k\lambda_1\int_{\text{LHP}} T\bar{T}-k\lambda_2\int_{\text{UHP}} T\bar{T}\big)+O(\lambda^2)}{\left(\int[\dd\phi]\mo(w_1)\mo^\dagger(w_2)e^{-\int_{\Sigma_1}\mathcal{L}_0[\phi]}\big(1-\lambda_1\int_{\text{LHP}} T\bar{T}-\lambda_2\int_{\text{LHP}} T\bar{T}\big)+O(\lambda^2)\right)^k}\nn\\
%=11&\frac{Z_k}{Z_1^k}\times\frac{\la\mo(w_1)\mo^\dagger(w_2)...\mo(w_{2k-1})\mo^\dagger(w_{2k})\ra_{\Sigma_k}}{\la\mo(w_1)\mo^\dagger(w_2)\ra^k_{\Sigma_1}}\nn\\
%&\times\left(1-\frac{k\lambda_1\int_{\text{LHP}}\dd^2 w\la T\bar{T}(w)\mo(w_1)...\mo^\dagger(w_{2k})\ra_{\Sigma_k}+k\lambda_2\int_{\text{UHP}}\dd^2 w\la T\bar{T}(w)\mo(w_1)...\mo^\dagger(w_{2k})\ra_{\Sigma_k}}{\la\mo(w_1)\mo^\dagger(w_2)...\mo(w_{2k-1})\mo^\dagger(w_{2k})\ra_{\Sigma_k}}\right)\nn\\
%&\times\left(1-\frac{\lambda_1\int_{\text{LHP}}\dd w\la T\bar{T}(w)\mo(w_1)\mo^\dagger(w_2)\ra_{\Sigma_1}+\lambda_2\int_{\text{UHP}}\dd w\la T\bar{T}(w)\mo(w_1)\mo^\dagger(w_2)\ra_{\Sigma_1}}{\la\mo(w_1)\mo^\dagger(w_2)\ra_{\Sigma_1}}\right)^{-k}+O(\lambda^2)\nn\\
=&\frac{Z_k}{Z_1^k}\times\frac{\la\mo(w_1)\mo^\dagger(w_2)...\mo(w_{2k-1})\mo^\dagger(w_{2k})\ra_{\Sigma_k}}{\la\mo(w_1)\mo^\dagger(w_2)\ra^k_{\Sigma_1}}\nn\\
&\times\Bigg(1-\frac{k\lambda_1\int_{\text{LHP}}\dd^2 w\la T\bar{T}(w)\mo(w_1)...\mo^\dagger(w_{2k})\ra_{\Sigma_k}+k\lambda_2\int_{\text{UHP}}\dd^2 w\la T\bar{T}(w)\mo(w_1)...\mo^\dagger(w_{2k})\ra_{\Sigma_k}}{\la\mo(w_1)\mo^\dagger(w_2)...\mo(w_{2k-1})\mo^\dagger(w_{2k})\ra_{\Sigma_k}}\nn\\
&~~+\frac{k\lambda_1\int_{\text{LHP}}\dd^2 w\la T\bar{T}(w)\mo(w_1)\mo^\dagger(w_2)\ra_{\Sigma_1}+k\lambda_2\int_{\text{UHP}}\dd^2 w\la T\bar{T}(w)\mo(w_1)\mo^\dagger(w_2)\ra_{\Sigma_1}}{\la\mo(w_1)\mo^\dagger(w_2)\ra_{\Sigma_1}}\Bigg)\label{ap1ta12}\\
&+\text{sub-leading~order terms}.\nn
\]
The two factors in the first line expression come from the vacuum contribution of the $k^{\rm th}$ R\'enyi entropy of subsystem $A$  (denoted as $S^{(k)}_{A,\text{vacuum}}$) and 
the operator $\mo$'s contribution of the $k^{\rm th}$ pseudo-R\'enyi entropy (denoted as $\Delta S^{(k)}_{A,0}$), respectively. Substituting \eqref{ap1ta12} into \eqref{pseudorenyi}, we obtain the $k^{\rm th}$ pseudo-R\'enyi entropy of subsystem $A$ under the $T\bar{T}$-deformation
\[
&\Delta S^{(k)}_{A,\lambda_1,\lambda_2}=S_{A,\lambda_1,\lambda_2}^{(k)}-S^{(k)}_{A,\text{vacuum}}-\Delta S^{(k)}_{A,0}\nn\\
=&\frac{1}{1-k}\log\Bigg(1-\frac{k\lambda_1\int_{\text{LHP}}\dd^2 w\la T\bar{T}(w)\mo(w_1)...\mo^\dagger(w_{2k})\ra_{\Sigma_k}+k\lambda_2\int_{\text{UHP}}\dd^2 w\la T\bar{T}(w)\mo(w_1)...\mo^\dagger(w_{2k})\ra_{\Sigma_k}}{\la\mo(w_1)\mo^\dagger(w_2)...\mo(w_{2k-1})\mo^\dagger(w_{2k})\ra_{\Sigma_k}}\nn\\
&~~+\frac{k\lambda_1\int_{\text{LHP}}\dd^2 w\la T\bar{T}(w)\mo(w_1)\mo^\dagger(w_2)\ra_{\Sigma_1}+k\lambda_2\int_{\text{UHP}}\dd^2 w\la T\bar{T}(w)\mo(w_1)\mo^\dagger(w_2)\ra_{\Sigma_1}}{\la\mo(w_1)\mo^\dagger(w_2)\ra_{\Sigma_1}}\Bigg)\nn\\
&+\text{sub-leading order terms}\nn\\
=&\frac{k}{k-1}\Bigg(\frac{\lambda_1\int_{\text{LHP}}\dd^2 w\la T\bar{T}(w)\mo(w_1)...\mo^\dagger(w_{2k})\ra_{\Sigma_k}+\lambda_2\int_{\text{UHP}}\dd^2 w\la T\bar{T}(w)\mo(w_1)...\mo^\dagger(w_{2k})\ra_{\Sigma_k}}{\la\mo(w_1)\mo^\dagger(w_2)...\mo(w_{2k-1})\mo^\dagger(w_{2k})\ra_{\Sigma_k}}\nn\\
&~~-\frac{\lambda_1\int_{\text{LHP}}\dd^2 w\la T\bar{T}(w)\mo(w_1)\mo^\dagger(w_2)\ra_{\Sigma_1}+\lambda_2\int_{\text{UHP}}\dd^2 w\la T\bar{T}(w)\mo(w_1)\mo^\dagger(w_2)\ra_{\Sigma_1}}{\la\mo(w_1)\mo^\dagger(w_2)\ra_{\Sigma_1}}\Bigg)\label{finalinapp1}\\
&+\text{sub-leading~order terms}.\nn
\]
Note that Eq. \eqref{finalinapp1} reduces to \eqref{equ4.2} when $\lambda_1=\lambda_2=\lambda$, and it reduces to \eqref{equ5.1} when $\mo=\mathbb{I}$, i.e., the identity operator.
\section{Details of integrals in pseudo-R\'enyi entropy with different deformed states}\label{appendixC}
In this appendix, we evaluate the integral of \eqref{equ5.3} in detail. The deformed pseudo-R\'enyi entropy according to appendix \ref{appendixB} can be evaluated as
\[
&\Delta S_{A,\lambda_1,\lambda_2}^{(k)}\nn\\
=&\frac{k\lambda_1}{k-1}\int_{\LHP\ \text{of}\ \Sigma_1} d^2 z\frac{1}{k^2\left|z\right|^{2k-2}}\nn\\
&\left(\frac{\left\la (T(z)+\frac{c(k^2-1)}{24}\frac{1}{z^2})(\bar{T}(\bar{z})+\frac{c(k^2-1)}{24}\frac{1}{\bar{z}^2})\mo_1(z_1,\bar{z}_1)...\mo^\dagger_{2k}(z_{2k},\bar{z}_{2k})\right\ra_{\Sigma_1}}{\la\mo_1(z_1,\bar{z}_1)...\mo^\dagger_{2k}(z_{2k},\bar{z}_{2k})\ra_{\Sigma_1}}\right.
-& \left.\frac{\la T\bar{T}(w,\bar{w})\mo_1\mo^\dagger_{2}\ra_{\Sigma_1}}{\la \mo_1\mo_{2}^\dagger\ra_{\Sigma_1}}\right)\nn\\
+&\frac{k\lambda_2}{k-1}\int_{\UHP\ \text{of} \ \Sigma_1} d^2 z\frac{1}{k^2\left|z\right|^{2k-2}}\nn\\
&\left(\frac{\left\la (T(z)+\frac{c(k^2-1)}{24}\frac{1}{z^2})(\bar{T}(\bar{z})+\frac{c(k^2-1)}{24}\frac{1}{\bar{z}^2})\mo_1(z_1,\bar{z}_1)...\mo^\dagger_{2k}(z_{2k},\bar{z}_{2k})\right\ra_{\Sigma_1}}{\la\mo_1(z_1,\bar{z}_1)...\mo^\dagger_{2k}(z_{2k},\bar{z}_{2k})\ra_{\Sigma_1}}\right.
-& \left.\frac{\la T\bar{T}(w,\bar{w})\mo_1\mo^\dagger_{2}\ra_{\Sigma_1}}{\la \mo_1\mo_{2}^\dagger\ra_{\Sigma_1}}\right).\label{appencix3.1}
\]
Similar to the discussion above, we still focus on the large c case. The leading order of the integral \eqref{appencix3.1} is simple, so we do not evaluate it in detail. We will calculate the integral \eqref{appencix3.1} of $\mo(c)$ in the rest of this appendix. At $\mo(c)$, using the conformal transformation \eqref{equ1.11} and Ward Identity, the integral \eqref{appencix3.1} can be rewritten as
\[
&\Delta S_{A,\lambda_1,\lambda_2}^{(k)}\nn\\
=&\frac{k\lambda_1}{k-1}\int_{\LHP\ \of\ \Sigma_1} d^2 z\frac{1}{k^2\left|z\right|^{2k-2}}\nn\\
&\left(\frac{c(k^2-1)}{24\bar{z}^2}\frac{\left\la T(z)\mo_1(z_1,\bar{z}_1)...\mo^\dagger_{2k}(z_{2k},\bar{z}_{2k})\right\ra_{\Sigma_1}}{\la\mo_1(z_1,\bar{z}_1)...\mo^\dagger_{2k}(z_{2k},\bar{z}_{2k})\ra_{\Sigma_1}}+\frac{c(k^2-1)}{24z^2}\frac{\left\la \bar{T}(\bar{z})\mo_1(z_1,\bar{z}_1)...\mo^\dagger_{2k}(z_{2k},\bar{z}_{2k})\right\ra_{\Sigma_1}}{\la\mo_1(z_1,\bar{z}_1)...\mo^\dagger_{2k}(z_{2k},\bar{z}_{2k})\ra_{\Sigma_1}}\right)\nn\\
+&\frac{k\lambda_2}{k-1}\int_{\UHP\ \of \ \Sigma_1} d^2 z\frac{1}{k^2\left|z\right|^{2k-2}}\nn\\
&\left(\frac{c(k^2-1)}{24\bar{z}^2}\frac{\left\la T(z)\mo_1(z_1,\bar{z}_1)...\mo^\dagger_{2k}(z_{2k},\bar{z}_{2k})\right\ra_{\Sigma_1}}{\la\mo_1(z_1,\bar{z}_1)...\mo^\dagger_{2k}(z_{2k},\bar{z}_{2k})\ra_{\Sigma_1}}+\frac{c(k^2-1)}{24z^2}\frac{\left\la \bar{T}(\bar{z})\mo_1(z_1,\bar{z}_1)...\mo^\dagger_{2k}(z_{2k},\bar{z}_{2k})\right\ra_{\Sigma_1}}{\la\mo_1(z_1,\bar{z}_1)...\mo^\dagger_{2k}(z_{2k},\bar{z}_{2k})\ra_{\Sigma_1}}\right)\nn\\
=&\frac{k\lambda_1}{k-1}\int_{\LHP\ \of\ \Sigma_1} d^2 z\frac{1}{k^2\left|z\right|^{2k-2}}\frac{c(k^2-1)}{24\la\mo_1(z_1,\bar{z}_1)...\mo^\dagger_{2k}(z_{2k},\bar{z}_{2k})\ra_{\Sigma_1}}\nn\\
&\left(\frac{1}{\bar{z}^2}\left(\sum\limits_{i=1}^{2k}\frac{h}{(z-z_i)^2}+\sum\limits_{j=0}^{k-2}\left(\frac{\partial_{2j+1}}{z-z_{2j+1}}+\frac{\partial_{2j+4}}{z-z_{2j+4}}\right)+\frac{\partial_{2}}{z-z_{2}}+\frac{\partial_{2k-1}}{z-z_{2k-1}}\right)\right.\nn\\
\times &\left\la \mo_1(z_1,\bar{z}_1)...\mo^\dagger_{2k}(z_{2k},\bar{z}_{2k})\right\ra_{\Sigma_1}\nn\\
&+\left.\frac{1}{z^2}\left(\sum\limits_{i=1}^{2k}\frac{\bar{h}}{(\bar{z}-\bar{z}_i)^2}+\sum\limits_{j=0}^{k-1}\left(\frac{\bar{\partial}_{2j+1}}{\bar{z}-\bar{z}_{2j+1}}+\frac{\bar{\partial}_{2j+2}}{\bar{z}-\bar{z}_{2j+2}}\right)\right)\left\la \bar{T}(\bar{z})\mo_1(z_1,\bar{z}_1)...\mo^\dagger_{2k}(z_{2k},\bar{z}_{2k})\right\ra_{\Sigma_1}\right)\nn\\
+&\frac{k\lambda_2}{k-1}\int_{\UHP\ \of \ \Sigma_1} d^2 z\frac{1}{k^2\left|z\right|^{2k-2}}\frac{c(k^2-1)}{24\la\mo_1(z_1,\bar{z}_1)...\mo^\dagger_{2k}(z_{2k},\bar{z}_{2k})\ra_{\Sigma_1}}\nn\\
&\left(\frac{1}{\bar{z}^2}\left(\sum\limits_{i=1}^{2k}\frac{h}{(z-z_i)^2}+\sum\limits_{j=0}^{k-2}\left(\frac{\partial_{2j+1}}{z-z_{2j+1}}+\frac{\partial_{2j+4}}{z-z_{2j+4}}\right)+\frac{\partial_{2}}{z-z_{2}}+\frac{\partial_{2k-1}}{z-z_{2k-1}}\right)\right.\nn\\
\times &\left\la \mo_1(z_1,\bar{z}_1)...\mo^\dagger_{2k}(z_{2k},\bar{z}_{2k})\right\ra_{\Sigma_1}\nn\\
&+\left.\frac{1}{z^2}\left(\sum\limits_{i=1}^{2k}\frac{\bar{h}}{(\bar{z}-\bar{z}_i)^2}+\sum\limits_{j=0}^{k-1}\left(\frac{\bar{\partial}_{2j+1}}{\bar{z}-\bar{z}_{2j+1}}+\frac{\bar{\partial}_{2j+2}}{\bar{z}-\bar{z}_{2j+2}}\right)\right)\left\la \bar{T}(\bar{z})\mo_1(z_1,\bar{z}_1)...\mo^\dagger_{2k}(z_{2k},\bar{z}_{2k})\right\ra_{\Sigma_1}\right).\label{appendix3.2}
\]
All the integrals can be fully evaluated after using the polar coordinate and taking proper regularization, and we will show the following integral as an example:
\[
&\int_{\LHP\ \of\ \Sigma_1} d^2 z \frac{z^2}{\left|z\right|^{2k+2}{(z-z_i)^2}}=\int_{0}^{\infty}\int_{-\pi}^0 d \rho d \theta \frac{e^{2 i \theta}}{\rho^{2k-1}(\rho e^{i\theta}-z_i)^2}\nn\\
=&\begin{cases}
    \int_{\epsilon}^{\left|z_i\right|}\rho^{-1-2k}\left(-\pi+\frac{2i\rho z_i}{\rho^2-z_i^2}+2i \Arctanh\left(\frac{z_i}{\rho}\right)\right)\ \ & z_i\in \LHP \ and\ \rho<\left|z_i\right|\\
    \int_{\left|z_i\right|}^{\infty}\rho^{-1-2k}\left(\pi+\frac{2i\rho z_i}{\rho^2-z_i^2}+2i \Arctanh\left(\frac{z_i}{\rho}\right)\right)\ \ & z_i\in \LHP\ and\ \rho>\left|z_i\right|\\
    \int_{\epsilon}^{\infty}\rho^{-1-2k}\left(\pi+\frac{2i\rho z_i}{\rho^2-z_i^2}+2i \Arctanh\left(\frac{z_i}{\rho}\right)\right)\ \ & z_i\in \UHP
\end{cases}\label{appendix3.3}
\]
where we have introduced $\epsilon$ as the cutoff. Integrate out $\rho$ in \eqref{appendix3.3} and regularize it through minimal subtraction, then we have
\[
\int_{\LHP\ \of\ \Sigma_1} d^2 z \frac{z^2}{\left|z\right|^{2k+2}{(z-z_i)^2}}=\begin{cases}
    \frac{\pi}{k\left|z_i\right|^{2k}}+\frac{\pi}{2k z_i^{2k}}+\frac{\pi}{z_i}  \ \ & z_i\in\LHP\\
    \frac{\pi}{2k z_i^{2k}}+\frac{\pi}{z_i} \ \ & z_i\in \UHP
\end{cases}.\label{appendix3.4}
\]
Other integrals in \eqref{appendix3.2} can be evaluated similarly, so we only list their results below. 
\[
\int_{\LHP\ \of\ \Sigma_1} d^2 z \frac{z^2}{\left|z\right|^{2k+2}{(z-z_i)}}=\begin{cases}
    \frac{z_i \pi}{k\left|z_i\right|^{2k}}-\frac{\pi}{2k z_i^{2k-1}}  \ \ & z_i\in \LHP\\
    -\frac{\pi}{2k z_i^{2k-1}} \ \ & z_i\in \UHP
\end{cases}.\label{appendix3.5}
\]
\[
\int_{\UHP\ \of\ \Sigma_1} d^2 z \frac{z^2}{\left|z\right|^{2k+2}{(z-z_i)^2}}=\begin{cases}
    \frac{\pi}{k\left|z_i\right|^{2k}}-\frac{\pi}{2k z_i^{2k}}-\frac{\pi}{z_i}  \ \ & z_i\in \UHP\\
    -\frac{\pi}{2k z_i^{2k}}-\frac{\pi}{z_i} \ \ & z_i\in \LHP
\end{cases}\label{appendix3.6}
\]
\[
\int_{\UHP\ \of\ \Sigma_1} d^2 z \frac{z^2}{\left|z\right|^{2k+2}{(z-z_i)}}=\begin{cases}
    \frac{z_i \pi}{k\left|z_i\right|^{2k}}+\frac{\pi}{2k z_i^{2k-1}}  \ \ & z_i\in \UHP\\
    \frac{\pi}{2k z_i^{2k-1}} \ \ & z_i\in \LHP
\end{cases}.\label{appendix3.7}
\]
\[
\int_{\UHP\ \of\ \Sigma_1} d^2 z \frac{\bar{z}^2}{\left|z\right|^{2k+2}{(\bar{z}-\bar{z}_i)^2}}=\begin{cases}
    \frac{\pi}{k\left|\bar{z}_i\right|^{2k}}+\frac{\pi}{2k \bar{z}_i^{2k}}+\frac{\pi}{\bar{z}_i}  \ \ & \bar{z}_i\in \LHP\\
    \frac{\pi}{2k \bar{z}_i^{2k}}+\frac{\pi}{\bar{z}_i} \ \ & \bar{z}_i\in \UHP
\end{cases}\label{appendix3.8}
\]
\[
\int_{\LHP\ \of\ \Sigma_1} d^2 z \frac{\bar{z}^2}{\left|z\right|^{2k+2}{(\bar{z}-\bar{z}_i)^2}}=\begin{cases}
    \frac{\pi}{k\left|\bar{z}_i\right|^{2k}}-\frac{\pi}{2k \bar{z}_i^{2k}}-\frac{\pi}{\bar{z}_i}  \ \ & \bar{z}_i\in \UHP\\
    -\frac{\pi}{2k \bar{z}_i^{2k}}-\frac{\pi}{\bar{z}_i} \ \ & \bar{z}_i\in \LHP
\end{cases}\label{appendix3.9}
\]
\[
\int_{\UHP\ \of\ \Sigma_1} d^2 z \frac{z^2}{\left|z\right|^{2k+2}{(z-z_i)}}=\begin{cases}
    \frac{\bar{z}_i \pi}{k\left|\bar{z}_i\right|^{2k}}+\frac{\pi}{2k \bar{z}_i^{2k-1}}  \ \ & \bar{z}_i\in \LHP\\
    \frac{\pi}{2k \bar{z}_i^{2k-1}} \ \ & \bar{z}_i\in \UHP
\end{cases}.\label{appendix3.10}
\]
\[
\int_{\LHP\ \of\ \Sigma_1} d^2 z \frac{z^2}{\left|z\right|^{2k+2}{(z-z_i)}}=\begin{cases}
    \frac{\bar{z}_i \pi}{k\left|\bar{z}_i\right|^{2k}}-\frac{\pi}{2k \bar{z}_i^{2k-1}}  \ \ & \bar{z}_i\in \UHP\\
    -\frac{\pi}{2k \bar{z}_i^{2k-1}} \ \ & \bar{z}_i\in \LHP
\end{cases}.\label{appendix3.11}
\]
Using results $\eqref{appendix3.4}$-$\eqref{appendix3.11}$ and $\eqref{appendix3.2}$, we get the correction to pseudo-R\'enyi entropy of primary operators with different $T\bar T$-deformation parameters
\[
&\Delta S_{A,\lambda_1,\lambda_2}^{(k)}\nn\\
=&\frac{c(k+1)}{24 k}\lambda_1\left[\sum\limits_{z_i \in \LHP}\frac{\pi h}{k \left|z_i\right|^{2k}}-\sum\limits_{z_{2i+1} \in  \LHP}\frac{2\pi h}{z_{2i+1}-z_{2i+4}}\frac{z_{2i+1}}{\left|z_{2i+1}\right|^{2k}}+\sum\limits_{z_{2i+4} \in  \LHP}\frac{2\pi h}{z_{2i+1}-z_{2i+4}}\frac{z_{2i+4}}{\left|z_{2i+4}\right|^{2k}}\right.\nn\\
&\sum\limits_{\bar z_i \in  \UHP}\frac{\pi h}{k \left|\bar z_i\right|^{2k}}-\sum\limits_{\bar z_{2i+1} \in \UHP}\frac{2\pi \bar  h}{\bar z_{2i+1}-\bar z_{2i+2}}\frac{\bar z_{2i+1}}{\left|\bar z_{2i+1}\right|^{2k}}+\sum\limits_{\bar z_{2i+2} \in  \UHP}\frac{2\pi \bar h}{\bar z_{2i+1}-\bar z_{2i+2}}\frac{\bar z_{2i+2}}{\left|\bar z_{2i+2}\right|^{2k}}\nn\\
+&\sum\limits_{j=1}^{2k}\left(\frac{\pi}{2k z_i^{2k}}+\frac{\pi}{z_i}\right)+\sum\limits_{j=0}^{k-2}\frac{\pi}{k}\left(\frac{-h}{z_{2j+1}-z_{2j+4}}\frac{z_{2j+1}}{z_{2j+1}^{2k}}+\frac{h}{z_{2j+1}-z_{2j+4}}\frac{z_{2j+4}}{z_{2j+4}^{2k}}\right)\nn\\
+&\frac{\pi}{k}\left(\frac{-h}{z_2-z_{2k-1}}\frac{z_2}{z_{2}^{2}}+\frac{h}{z_2-z_{2k-1}}\frac{z_{2k-1}}{z_{2k-1}^{2}}\right)-\sum\limits_{j=1}^{2k}\left(\frac{\pi}{2k \bar z_i^{2k}}+\frac{\pi}{\bar z_i}\right)\nn\\
-&\left.\sum\limits_{j=0}^{k-1}\frac{\pi}{k}\left(\frac{-\bar h}{\bar z_{2j+1}-\bar z_{2j+2}}\frac{\bar z_{2j+1}}{\bar z_{2j+1}^{2k}}+\frac{\bar h}{\bar z_{2j+1}-\bar z_{2j+2}}\frac{\bar z_{2j+2}}{\bar z_{2j+2}^{2k}}\right)\right] \nn\\
+&\frac{c(k+1)}{24 k}\lambda_2\left[\sum\limits_{z_i \in \UHP}\frac{\pi h}{k \left|z_i\right|^{2k}}-\sum\limits_{z_{2i+1} \in  \UHP}\frac{2\pi h}{z_{2i+1}-z_{2i+4}}\frac{z_{2i+1}}{\left|z_{2i+1}\right|^{2k}}+\sum\limits_{z_{2i+4} \in  \UHP}\frac{2\pi h}{z_{2i+1}-z_{2i+4}}\frac{z_{2i+4}}{\left|z_{2i+4}\right|^{2k}}\right.\nn\\
&\sum\limits_{\bar z_i \in  \LHP}\frac{\pi h}{k \left|\bar z_i\right|^{2k}}-\sum\limits_{\bar z_{2i+1} \in \LHP}\frac{2\pi \bar h}{\bar z_{2i+1}-\bar z_{2i+2}}\frac{\bar z_{2i+1}}{\left|\bar z_{2i+1}\right|^{2k}}+\sum\limits_{\bar z_{2i+2} \in  \LHP}\frac{2\pi \bar h}{\bar z_{2i+1}-\bar z_{2i+2}}\frac{\bar z_{2i+2}}{\left|\bar z_{2i+2}\right|^{2k}}\nn\\
-&\sum\limits_{j=1}^{2k}\left(\frac{\pi}{2k z_i^{2k}}+\frac{\pi}{z_i}\right)-\sum\limits_{j=0}^{k-2}\frac{\pi}{k}\left(\frac{-h}{z_{2j+1}-z_{2j+4}}\frac{z_{2j+1}}{z_{2j+1}^{2k}}+\frac{h}{z_{2j+1}-z_{2j+4}}\frac{z_{2j+4}}{z_{2j+4}^{2k}}\right)\nn\\
-&\frac{\pi}{k}\left(\frac{-h}{z_2-z_{2k-1}}\frac{z_2}{z_{2}^{2}}+\frac{h}{z_2-z_{2k-1}}\frac{z_{2k-1}}{z_{2k-1}^{2}}\right)+\sum\limits_{j=1}^{2k}\left(\frac{\pi}{2k \bar z_i^{2k}}+\frac{\pi}{\bar z_i}\right)\nn\\
+&\left.\sum\limits_{j=0}^{k-1}\frac{\pi}{k}\left(\frac{-\bar h}{\bar z_{2j+1}-\bar z_{2j+2}}\frac{\bar z_{2j+1}}{\bar z_{2j+1}^{2k}}+\frac{\bar h}{\bar z_{2j+1}-\bar z_{2j+2}}\frac{\bar z_{2j+2}}{\bar z_{2j+2}^{2k}}\right)\right]
\label{appendix3.12}\]
Replacing all the coordinates in \eqref{appendix3.12} with \eqref{A=INFzcoordin}, we get the  correction to the $k^{\rm th}$ pseudo-R\'enyi entropy with different deformed parameters under $T\bar T$ deformation at the late time
\[
&\Delta S_{A,\lambda_1,\lambda_2}^{(k)}=\nn\\
&\frac{c(k+1)}{24 k^2}\lambda_1\Bigg\{\frac{\left((-t+x_1)^{\frac{1}{k}}(t-x_1)^{\frac{1}{k}}-e^{\frac{2\pi i}{k}}(-t-x_2)^{\frac{1}{k}}(t-x_2)^{\frac{1}{k}}\right)(x_1^2-x_2^2)}{(t^2-x_1^2)\left((-t-x_1)^{\frac{1}{k}}-e^{\frac{2\pi i}{k}}(-t-x_2)^{\frac{1}{k}}\right)\left((t-x_1)^{\frac{1}{k}}-(t-x_2)^{\frac{1}{k}}\right)(t^2-x_2^2)}\nn\\
+&\frac{1}{2} \pi \Bigg [\frac{\left(x_1-x_2\right) \left(2 t-x_1-x_2\right) \left(\left(t-x_1\right){}^{1/k}+\left(t-x_2\right){}^{1/k}\right)}{\left(t-x_1\right){}^2 \left(t-x_2\right){}^2 \left(\left(t-x_1\right){}^{1/k}-\left(t-x_2\right){}^{1/k}\right)}+\frac{2 \left(\left(-t-x_1\right){}^{\frac{1}{k}-2}-e^{\frac{2 i \pi }{k}} \left(-t-x_2\right){}^{\frac{1}{k}-2}\right)}{-\left(-t-x_1\right){}^{1/k}+e^{\frac{2 i \pi }{k}} \left(-t-x_2\right){}^{1/k}}\nn\\
+&2 t^2+2 t \left(x_1+x_2\right)  +2 \left(\frac{1}{t+x_2}+\frac{1}{t+x_1}\right)+x_1^2+x_2^2\Bigg ]\Bigg\}
+\mo(\epsilon,c^0,\lambda_1^2)\nn\\
+&\frac{c(k+1)}{24 k^2}\lambda_2\Bigg \{\frac{\left((-t+x_1)^{\frac{1}{k}}(t-x_1)^{\frac{1}{k}}-e^{\frac{2\pi i}{k}}(-t-x_2)^{\frac{1}{k}}(t-x_2)^{\frac{1}{k}}\right)(x_1^2-x_2^2)}{(t^2-x_1^2)\left((-t-x_1)^{\frac{1}{k}}-e^{\frac{2\pi i}{k}}(-t-x_2)^{\frac{1}{k}}\right)\left((t-x_1)^{\frac{1}{k}}-(t-x_2)^{\frac{1}{k}}\right)(t^2-x_2^2)}\nn\\
-&\frac{1}{2} \pi \Bigg [\frac{\left(x_1-x_2\right) \left(2 t-x_1-x_2\right) \left(\left(t-x_1\right){}^{1/k}+\left(t-x_2\right){}^{1/k}\right)}{\left(t-x_1\right){}^2 \left(t-x_2\right){}^2 \left(\left(t-x_1\right){}^{1/k}-\left(t-x_2\right){}^{1/k}\right)}+\frac{2 \left(\left(-t-x_1\right){}^{\frac{1}{k}-2}-e^{\frac{2 i \pi }{k}} \left(-t-x_2\right){}^{\frac{1}{k}-2}\right)}{-\left(-t-x_1\right){}^{1/k}+e^{\frac{2 i \pi }{k}} \left(-t-x_2\right){}^{1/k}}\nn\\
+&2 t^2+2 t \left(x_1+x_2\right)  +2 \left(\frac{1}{t+x_2}+\frac{1}{t+x_1}\right)+x_1^2+x_2^2\Bigg ]\Bigg\}+\mo(\epsilon,c^0,\lambda_2^2).
\]
\bibliographystyle{JHEP}
\bibliography{TTbarREF}
\end{document}